\documentclass[12pt, draftclsnofoot, onecolumn]{IEEEtran}

\usepackage{amssymb}
\usepackage{graphicx}
\usepackage{multirow}
\usepackage{epstopdf}
\usepackage{subfigure}
\usepackage{amsmath}
\usepackage{url}
\usepackage{float}
\usepackage{colortbl}
\usepackage{makecell}
\usepackage{algorithm}
\usepackage{algorithmic}
\usepackage[table]{xcolor}

\begin{document}
\title{\huge{DeepMA: End-to-end Deep Multiple Access for Wireless Image Transmission in Semantic Communication}}
\author{Wenyu Zhang, Kaiyuan Bai, Sherali Zeadally, Haijun Zhang, Hua Shao, Hui Ma, Victor C. M. Leung
\thanks{Wenyu Zhang, Hua Shao, and Hui Ma are with the School of Intelligence Science and Technology, and the Institute of Artificial Intelligence, University of Science and Technology Beijing, Beijing, 100083, China. Emails: wyzhang@ustb.edu.cn, shaohua@ustb.edu.cn, hui\_ma@ustb.edu.cn.}
\thanks{Kaiyuan Bai is with the China Telecom Research Institute, Beijing, 100096, China. Email: baiky1@chinatelecom.cn}
\thanks{Sherali Zeadally is with the College of Communication and Information, University of Kentucky, Lexington, KY 40506 USA. Email: szeadally@uky.edu}
\thanks{Haijun Zhang is with the Beijing Advanced Innovation Center for Materials Genome Engineering, Beijing Engineering and Technology Research Center for Convergence Networks and Ubiquitous Services, University of Science and Technology Beijing, Beijing 100083, China. E-mail: haijunzhang@ieee.org.}
\thanks{Victor C. M. Leung is with the College of Computer Science and Software Engineering, Shenzhen University, Shenzhen 518060, China, and also with the Department of Electrical and Computer Engineering, The University of British Columbia, Vancouver, BC V6T 1Z4, Canada E-mail: vleung@ieee.org.}}

\maketitle
\begin{abstract}
Semantic communication is a new paradigm that exploits deep learning models to enable end-to-end communications processes, and recent studies have shown that it can achieve better noise resiliency compared with traditional communication schemes in a low signal-to-noise (SNR) regime. To achieve multiple access in semantic communication, we propose a deep learning-based multiple access (DeepMA) method by training semantic communication models with the abilities of joint source-channel coding (JSCC) and orthogonal signal modulation. DeepMA is achieved by a DeepMA network (DMANet), which is comprised of several independent encoder-decoder pairs (EDPs), and the DeepMA encoders can encode the input data as mutually orthogonal semantic symbol vectors (SSVs) such that the DeepMA decoders can detect and recover their own target data from a received mixed SSV (MSSV) superposed by multiple SSV components transmitted from different encoders. We describe frameworks of DeepMA in wireless device-to-device (D2D), downlink, and uplink channel multiplexing scenarios, along with the training algorithm. We evaluate the performance of the proposed DeepMA in wireless image transmission tasks and compare its performance with the attention module-based deep JSCC (ADJSCC) method and conventional communication schemes using better portable graphics (BPG) and Low-density parity-check code (LDPC). The results obtained show that the proposed DeepMA can achieve effective, flexible, and privacy-preserving channel multiplexing process, and demonstrate that our proposed DeepMA approach can yield comparable bandwidth efficiency compared with conventional multiple access schemes.
\end{abstract}


\section{Introduction}

In a conventional communication system, the data processing process mainly includes the following three parts: source coding for improving transmission efficiency by reducing the information redundancy of the source data, channel coding for enhancing the transmitting reliability by using error-check or error-correcting coding techniques, and signal modulation for improving the channel resource utilization efficiency by enabling multiple users to transmit data over a shared physical channel, while transforming the digital signals as high-frequency waveforms that are suitable for wireless transmission \cite{tse2005fundamentals}. These processes are conducted by specific and separable functional blocks, which are designed independently without jointly considering the influences of other blocks. This separable architecture significantly reduces the design, development, and maintenance costs of the communication system, but it is suboptimal from the perspective of end-to-end optimization \cite{liu2022high}\cite{qin2021semantic}.

In recent years, the rapid development of artificial intelligence (AI), especially deep learning (DL)-based autoencoder models, has enabled us to achieve the end-to-end semantic communication process by leveraging the powerful data compression and noise resiliency capability of deep learning models. A typical semantic communication system (SCS) is mainly composed of a trainable semantic encoder, a non-trainable noisy physical communication channel, and a trainable semantic decoder \cite{qin2021semantic}\cite{kountouris2021semantics}. Both semantic encoder and semantic decoder are realized by using deep learning models, such as convolution neural networks (CNNs), recurrent neural networks (RNN), and attention networks. {\color{black}A typical semantic encoder automatically conducts the joint source-channel coding process to extract the semantic features that are suitable for transmission over a noisy physical channel. The feature data contains the semantic contents and channel code of the input data, and in the physical channel it can be corrupted with noises and interferences after the transmission process, which may cause semantic distortion to the final reconstructed result. In accordance with the semantic encoder, a paired semantic decoder can automatically complete the inference task directly from the received semantic symbols corrupted by noises or interferences. If the inference task is recovering the original input data, then the decoder must perform the channel decoding and source decoding process jointly.}

Recent studies \cite{qin2021semantic}\cite{bourtsoulatze2019deep}\cite{weng2021semantic}\cite{xie2021deep} have shown that semantic communication can achieve high communication efficiency and noise resiliency in various data transmission scenarios, such as video/image transmission \cite{bourtsoulatze2019deep}\cite{tung2022deepwive}, speech transmission \cite{weng2021semantic}, and text transmission \cite{xie2021deep}. More specifically, by utilizing the powerful feature extraction capability of deep learning models, semantic communication models can achieve a higher data compression ratio compared with classical data compression methods so that the amount of transmitted data can be further reduced. Second, deep learning models can be trained with powerful noise resiliency capability, which enables the semantic communication models to complete the inference task to obtain high-quality results when there is some noise and interferences. Semantic encoders and decoders are jointly trained end-to-end between the source input data at the transmitter and the final inference result at the receiver. Thus, the source channel encoding and decoding process are jointly optimized, and it is globally optimal compared with conventional communication systems with separate functional blocks. Semantic communication can achieve a relatively stronger noise resiliency in a low channel signal to noise ratio (SNR) regime, while in conventional separable communication schemes, the transmission reliability suffers from the 'cliff effect', i.e., the transmission performance drops sharply when the channel condition is poor, and even fails to transmit data.

In conventional communications, multiple access (multi-access) is a commonly used technology that enables multiple users to multiplex the communication channel without interfering with each other, which is helpful for improving the utilization efficiency of the channel resources, and typical methodologies include orthogonal frequency division multiple access (OFDMA) \cite{myung2006single}, time division multiple access (TDMA) \cite{ergen2010tdma}, code division multiple access (CDMA) \cite{gilhousen1991capacity}, and non-orthogonal multiple access (NOMA) \cite{dai2018survey}. Among them, OFDMA achieves multiplexing by allocating the dedicated frequency sources to the user equipments (UEs), and TDMA avoids interference by dividing and allocating UEs with different transmission timeslot resources. NOMA can support the co-frequency transmission by using a successive interference cancellation technique. CDMA is an orthogonal multiple access (OMA) method that can achieve channel multiplexing by modulating the data bits of multiple users using a set of orthogonal codes, such that the data can be transmitted over with the same frequency, and the receiver can recover the data by demodulating the received data with the same code. CDMA enables multiple UEs to transmit data over a shared physical channel with the same frequency without worrying about collisions while protecting the data privacy when the basis code used is unknown to other UEs, and it has been widely used in conventional communication systems \cite{tse2005fundamentals}.

{\color{black}
As a new communication paradigm, currently, semantic communication research mainly addresses the deep learning-based joint source-channel coding (DeepJSCC) problem for point-to-point communication tasks \cite{qin2021semantic} but how to achieve co-frequency multiple access to enable multiple semantic data streams to be transmitted simultaneously with the same frequency remains an unexplored problem. Compared with conventional multiplexing methods, realizing co-frequency multi-access in semantic communication has the following differences:
\begin{itemize}
  \item \textbf{End-to-end vs separable}: Semantic communication achieves end-to-end communication by leveraging the powerful learning capability of deep learning models. If we can achieve DeepJSCC and orthogonal modulation in one OMA-enabled semantic communication model, then there is no need to use the conventional separable OFDMA or TDMA-based multiple access methods because the semantic symbols are already orthogonal to each other, and the system complexity will be significantly reduced.
  \item \textbf{Continuous vs binary}: Conventional CDMA is only suitable for transmitting bit data, and it is not suitable for semantic codes with continuous values. Moreover, from the perspective of bandwidth efficiency, CDMA is not an ideal choice because its spread spectrum process will significantly decrease the bandwidth efficiency. Therefore, it is necessary to develop a specific multi-access method that is suitable for transmitting continuous symbols in semantic communication systems with high bandwidth efficiency.
  \item \textbf{Lossy vs lossless}: Conventional communication requires that the transmitted bit data be lossless, i.e., every bit is correctly transmitted, while semantic communication is lossy and the received symbols are allowed to be misinterpreted or distorted to the original transmitted symbols. For CDMA, when the received codes has error bits, the transmission fails, which is not compatible to the lossy semantic communication process.
\end{itemize}

Thus, we can see that conventional multiple access methods are not the best choices for semantic communication systems. In particular, conventional CDMA is actually not applicable in semantic communication due to the differences we have mentioned above.
To achieve co-frequency OMA in semantic communication, we propose deep learning-based multiple access (DeepMA) to enable lossy and orthogonal channel multiplexing by exploiting the capabilities of the DL models. We summarize the main contributions of this work as follows:
}

\begin{itemize}
\item We propose a channel multiplexing-enabled DeepMA network (DMANet), which is comprised of multiple DeepMA encoder-decoder pairs (EDP), and the EDPs are trained with the abilities of JSCC and orthogonal signal modulation. Similar to CDMA, a DMANet can support channel multiplexing by transmitting multiple orthogonal semantic symbol vectors (SSVs) simultaneously, and the DeepMA decoders can recover the target data from a received mixed SSV (MSSV) superposed by several different SSVs. In {\color{black}contrast} to CDMA, the orthogonal modulation process in DeepMA is learned by the model training process, and it is automatically conducted without the help of an orthogonal basis. {\color{black}DeepMA can achieve end-to-end and lossy semantic communication for continuous semantic symbol data with channel multiplexing. In addition, to avoid unnecessary computation cost, a SSV gate (SSVG) is introduced to conduct the user detection process to avoid the unnecessary decoding process of undesired data.}
\item We conducted simulations to evaluate the performance of the proposed DeepMA in wireless image transmission tasks. First, we developed an attention module and residual structure-based DeepMA encoder and decoder models, and then trained the models by using CIFAR10 and ImageNet data. After training, we compare the peak-signal-to-noise (PSNR) performances of the proposed DeepMA models, ADJSCC models, and conventional communications schemes that use better portable graphics (BPG) \cite{BPG} for source coding and low-density parity-check (LDPC) for channel coding \cite{liu2022high}. We used the ADJSCC model as the reference method for non-multiplexing semantic transmission to show how the multiple access process influences the transmission capacity of DeepMA. We used the conventional multiple access schemes as baselines to demonstrate the performance of our proposed DeepMA. To demonstrate the performance of the proposed DeepMA, we first present some illustrative examples to show the mutual orthogonality of the SSVs, which is also helpful for us to understand why DeepMA can achieve OMA, and how to use SSV orthogonality to conduct user detection. The illustrative examples show that a DeepMA decoder can effectively recover the target image from an MSSV that contains the SSV transmitted from the paired DeepMA encoder but cannot obtain any useful information if the paired DeepMA encoder does not send the SSV, showing that the proposed DeepMA can achieve flexible and privacy-protected multiple access. We compare the (PSNR) performances achieved by DeepMA, ADJSCC, and conventional BPG+LDPC schemes with different channel SNR, and the results demonstrate that the proposed can achieve high-quality recovered images with channel multiplexing, and show that DeepMA can achieve better noise-resiliency compared with conventional BPG+LDPC schemes in a very low SNR regime.
\end{itemize}

We organize the remainder of this paper as follows. Section 2 discusses related works. Section 3 presents our proposed DeepMA method along with the training algorithm. Section 4 presents the performance evaluation results on wireless image transmissions. We also describe the architecture of the DMANet used. Finally, Section 5 concludes the paper.

\emph{Notation}: $\mathbb{C}^n$ and $\mathbb{R}^n$ {\color{black}mean} the complex and real data with total dimensions $n$ respectively. $x\sim \mathcal{CN}(\mu_x, \sigma_x^2)$ designates variable $x$ follows a circularly-symmetric complex Gaussian distribution with mean $\mu_x$ and $\sigma_x^2$. Variables with bold fonts mean that they are matrices or vectors. $\mathbf{a}\cdot \mathbf{b}$ means the dot-product of vectors $\mathbf{a}$ and $\mathbf{b}$. $\mathbf{x}^*$ means the {\color{black}conjugate} transpose of complex variable $\mathbf{x}$. Finally, $\mathbb{E}[x]$ means the expectation value of variable $x$, and $\mathbf{I}$ denotes the identity matrix.

\section{Related Work}

Semantic communication can be used for dealing with a broad range of intelligent inference tasks over the network \cite{qin2021semantic}\cite{shi2021semantic}. These tasks include answering visual questions \cite{xie2021task}, image retrieval \cite{jankowski2020wireless}, speech recognition \cite{weng2021semantic}, and video classification \cite{ozyilkan2020deep}.
In this paper, we mainly consider the semantic communication scenario with the goal of recovering the input data at the receiver, and the data modalities considered can be image/video, speech audio, and language text. For example, for text data transmission, Farsad et.al {\color{black}\cite{farsad2018deep}} proposed a bidirectional long short-term memory (BLSTM) network to achieve JSCC. The proposed method achieved a lower word error rate compared with the traditional gzip and Huffman coding methods. Xie et.al \cite{xie2021deep} proposed a deep learning-based semantic communication (DeepSC) system that is built by the Transformer based semantic encoder/decoder along with a fully connected layer-based channel encoder/decoder. During the training process, DeepSC maximizes sentence similarity  by using the bilingual evaluation understudy (BLEU) score, and transfer learning is used for enabling the system to be applicable to different communication environments. Experimental results obtained demonstrate that the proposed SCS approach can achieve higher BLEU compared with traditional communication systems, especially in situations where the channel SNR is low. In a similar way, the authors of \cite{weng2021semantic} proposed DeepSC for speech (DeepSC-S) transmission, and the results demonstrated that DeepSC-S outperforms the traditional speech transmission schemes for both speech recovery quality and noise-resiliency.

In this paper, we focus on using semantic communication for solving wireless image transmission tasks, and related works have shown that DeepJSCC methods can achieve excellent data compression and noise resiliency abilities \cite{cui2021asymmetric}\cite{sujitha2021optimal}\cite{mentzer2020high}\cite{tung2022deepjscc}. In \cite{bourtsoulatze2019deep}, the authors proposed a DeepJSCC method for wireless image transmission. Simulation results showed that the proposed DeepJSCC outperforms the traditional JPEG and JPEG-2000 image compression methods in low SNR regimes. Compared with traditional digital communication schemes, DeepJSCC does not suffer from the 'cliff effect', and it has an strong noise resiliency capability against the degradation of channel SNR. To further improve the performance of JSCC, Kurka et.al \cite{kurka2020deepjscc} proposed the use of channel output feedback to enhance the reliability of the transmission process. The proposed method can achieve better image reconstruction quality, but it requires multiple transmissions for one input data. Thus, the communication efficiency will be reduced. Later, the authors proposed the successive refinement-based DeepJSCC method to enhance the quality of the reconstructed images at the receiver \cite{kurka2021bandwidth}. From the perspective of ensemble learning \cite{sagi2018ensemble}, the performance improvements achieved by the above two methods comes from the integration of multiple DeepJSCC models with diverse DeepJSCC models and physical channels, and the quality of the final combined results can be improved compared with using only one single model \cite{yang2020image}\cite{su2020ensemble}.

The proposed DeepJSCC models above suffer from the SNR adaptation problem, i.e., a DeepJSCC model can only perform well when the channel SNR is similar to its training SNR. To achieve the best performance, we need to train several different DeepJSCC models to cover both low and high SNR regimes. To solve this problem, Xu et.al \cite{xu2021wireless} proposed an attention-based DeepJSCC (ADJSCC), which can adaptively adjust the joint source-channel coding process according to the variation of channel SNR. In a low SNR regime, the attention module can automatically guide the model to assign more coding resources to the channel coding process to enhance transmission reliability. On the other hand, in a high SNR regime, more encoding resources will be assigned to the source coding process to improve the compression ratio. Simulation results demonstrated that one ADJSCC model can achieve the best achievable performance of several DeepJSCC models training with different channel SNRs combined. Therefore, ADJSCC can greatly reduce the training and deployment cost of Deep learning-based JSCC models.

Except for the SNR adaptation problem, previous DeepJSCC models were trained with specific code rates or compression ratios and cannot be changed after training. To achieve multi-rate transmission, in \cite{yang2022deep}, the authors proposed an adaptive DeepJSCC model that can automatically adjust its rate using a single network model. More specifically, in the proposed method the encoder is divided into two separate models: source encoder and channel encoder, and the semantic symbols obtained are divided into two parts: the non-selective part that is always active for transmission, and the selective part that can be active or inactive for transmission. A policy network is introduced for determining which selective channel codes are active for transmission according to the image contents and channel condition. In this way, the proposed method can support multi-rate transmission for semantic communication. In the above work, the proposed method cannot achieve an arbitrary transmission rate, and transmission quality cannot be guaranteed. In \cite{zhang2023predictive}, we proposed a predictive and adaptive deep coding (PADC) framework, which can flexibly adjust compression rates at the same time ensuring the transmission quality constraint for every single image thereby providing an effective solution for solving the rate adaptation problem in DeepJSCC.

All the works discussed above focus on the DeepJSCC problem for achieving point-to-point semantic communication scenario, and the OMA problem has not been investigated. As such, in this paper, we propose DeepMA to achieve channel multiplexing and demonstrate that, by leveraging deep learning, we can obtain effective OMA in semantic communication, and can reap better bandwidth efficiency compared with conventional multiple access schemes in low SNR regimes.

\begin{figure}
\centering
\includegraphics[width=0.75\textwidth]{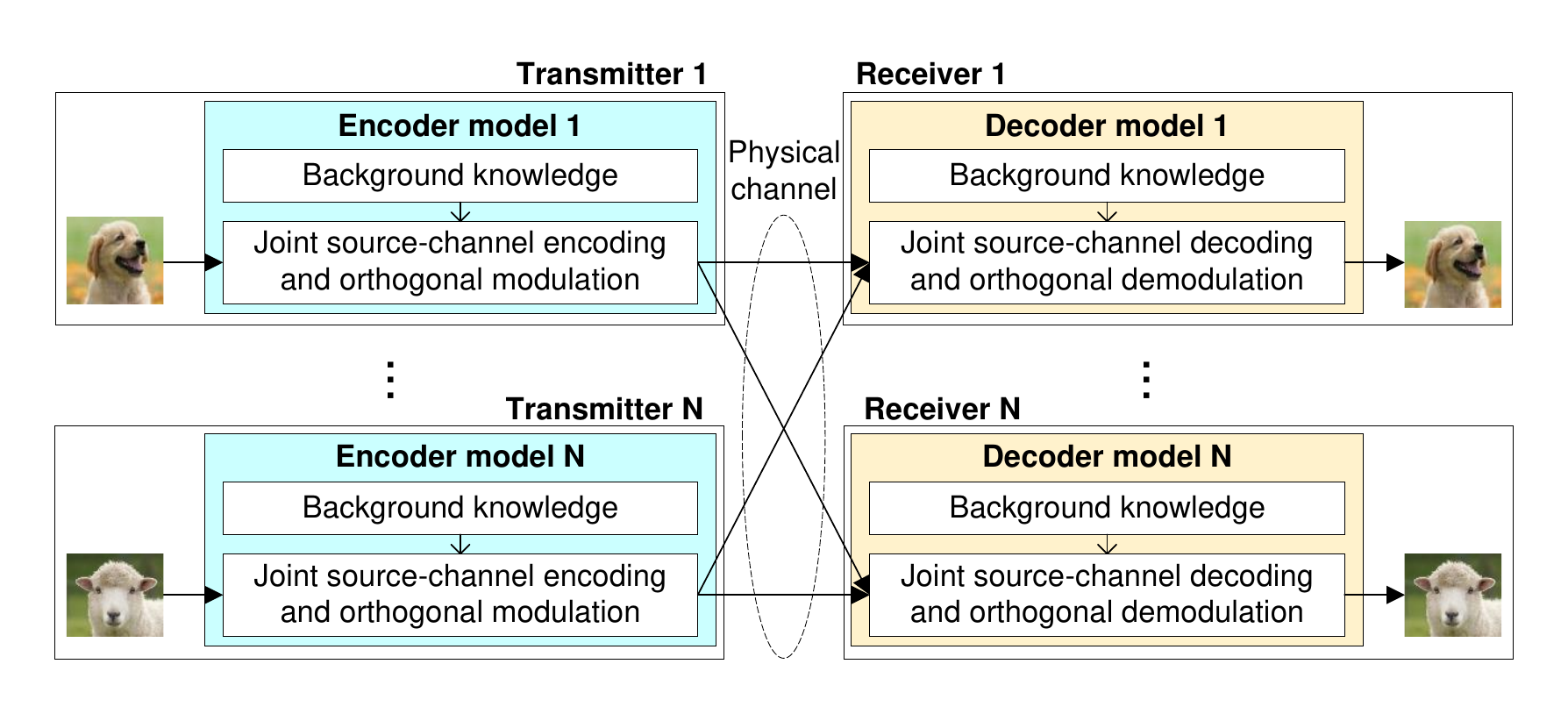}
\caption{{\color{black}The proposed DeepMA framework.}}
\label{sys_mod}
\end{figure}

\begin{figure*}
\centering
\includegraphics[width=0.98\textwidth]{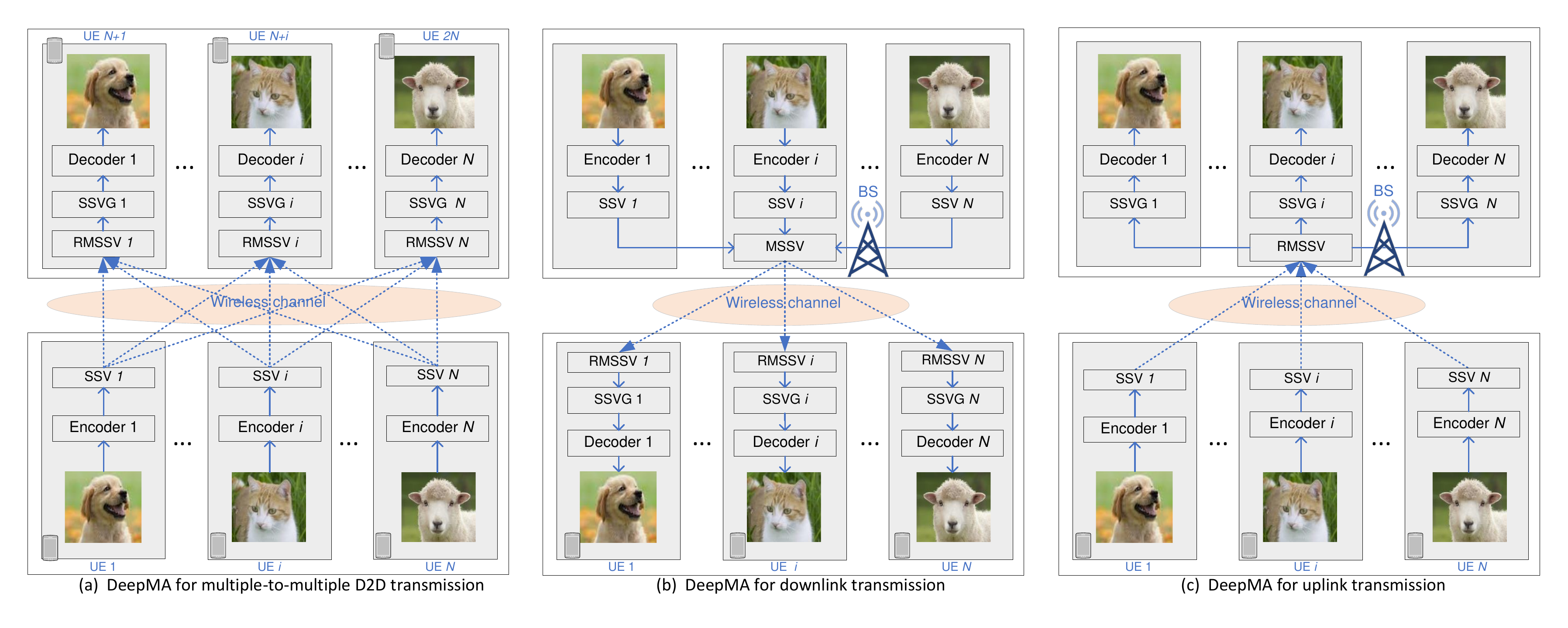}
\caption{{\color{black}DeepMA for wireless D2D, downlink, and uplink transmissions with channel multiplexing.}}
\label{DeepMA}
\end{figure*}

\section{The Proposed DeepMA}
{\color{black}
Fig.\ref{sys_mod} shows the basic framework of DeepMA, wherein the deep coding and modulation (DCM) processes of a transmitter and a receiver are realized by a semantic encoder and semantic decoder respectively. More specifically, for the encoder, except for source encoding and channel encoding, it also conducts orthogonal signal modulation such that the transmitted SSVs are mutually orthogonal. Accordingly, the decoder conducts the joint orthogonal signal demodulation, channel decoding and source decoding processes, such that it can detect and recover the transmitted data from several superposed SSVs. The encoders and decoders obtain the DCM abilities from a model training process, in which the EDPs can automatically learn background knowledge on how to achieve OMA with different training data and channel conditions. Given different training data and channel environments, the background knowledge learned can be different. The transmission quality of DeepMA is jointly influenced by factors such as semantic ambiguity, model capability, data compression ratio, and physical channel. In particular, in DeepMA the transmitted SSVs are mixed at the physical channel, and its main goal is to achieve OMA to eliminate the influence of other SSVs transmitted from unpaired encoders. In this section, we will describe the system model, design, and training algorithm of DeepMA.


\subsection{System Model}

In addition to Fig. \ref{sys_mod}, Fig. \ref{DeepMA} shows the system model of the proposed DeepMA in more detail. We denote the set of transceivers as $\mathcal{V} = \left\{(\text{Tx}_1, \text{Rx}_1), \ldots, (\text{Tx}_N, \text{Rx}_N)\right\}$, in which $\text{Tx}_i$ and $\text{Rx}_i$ denote the transmitter and receiver of transceiver $i$ respectively. We construct a DeepMA network (DMANet) that is composed of $N$ DeepMA encoder-decoder pairs (EDPs), denoted as $\mathcal{P} = \left\{(E_{\phi_1},D_{\theta_1}),\ldots, (E_{\phi_N},D_{\theta_N}) \right\}$, in which $E_{\phi_i}$ means the DeepMA encoder model of EDP $i$ is parameterized by $\phi_i$, and $D_{\theta_i}$ designates the corresponding DeepMA decoder network parameterized by $\theta_i$. The encoder model $E_{\phi_i}$ and decoder model $D_{\theta_i}$ are deployed at the transmitter $\text{Tx}_i$ and receiver $\text{Rx}_i$ respectively. When $N=1$, the DMANet is equivalent to a DeepJSCC network. The deployment of DeepMA model can be flexible, including the following three scenarios:
\begin{itemize}
  \item \textbf{Device-to-device (D2D)}: Each UE is equipped with one encoder model or decoder model, and each transmitter or receiver corresponds to one UE.
  \item \textbf{Downlink}: All encoders are deployed at the base station (BS), and each UE receiver has a decoder model. Therefore, we have only one transmitter, but multiple receivers.
  \item \textbf{Uplink}: All the decoders are deployed at the BS, and each UE transmitter has an encoder model. Therefore, we have only one receiver, but multiple transmitters.
\end{itemize}

We denote $\mathbf{z}_i \in \mathbb{C}^{K}$ as the SSV transmitted by $\text{Tx}_i$, and each symbol of SSV $\mathbf{z}_i$ experiences an independent transmission process over through a noisy physical channel. We consider the AWGN channel and slow Rayleigh fading channel in this paper, which are two commonly used channel models. In a co-frequency multi-access scenario, the received signal of $\text{Rx}_i$ in AWGN channel case is formulated as:
\begin{equation}
\tilde{\mathbf{z}_i} = \sum_{i=1}^{N} \mathbf{z}_i + \mathbf{n}_i,
\end{equation}
where $\mathbf{n}_i \in \mathcal{CN}(0, \sigma_i^2 \mathbf{I})$ denotes the independent identically distributed (i.i.d) AWGN samples with power $\sigma_i^2$. In the slow Rayleigh fading channel case, the transmitted signals experience independent channel fading processes. We equip each node with a single antenna, then the channel state information (CSI) among the transmitters and receivers is denoted by a matrix $\mathbf{H} \in \mathbb{C}^{N\times N}$, in which $h_{i,j}$ denotes the CSI from transmitter $\text{Tx}_i$ and receiver $\text{Rx}_j$. For transceiver $i$, the CSI from its transmitter to receiver $h_{i,i}$ must be known to the receiver, but other CSI information is not required to be known to the receiver.

In the D2D transmission scenario, as Fig.\ref{DeepMA}(a) shows, there are $N$ UEs transmitting SSVs, and accordingly there are $N$ UEs receiving data, and one UE transmitter has one and only one paired UE receiver. For SSV $\mathbf{z}_i$, it is transmitted to the corresponding receiver, at the same time superposed with other SSVs, and corrupted with AWGN, and the mixed SSV (MSSV) received is:

\begin{equation}\label{eq:fad_channel1}
\mathbf{z}_i^{\text{rev}}  = \sum_{k=1}^{N}  h_{k,i} \mathbf{z}_k + \mathbf{n}_i.
\end{equation}

In the downlink transmission scenario, as Fig.\ref{DeepMA}(b) shows, in the BS, the data arrived is first encoded as the SSVs by using the DeepMA encoders, then superposed as the MSSV, and subsequently transmitted to the UEs through the downlink channel. In this scenario, we can simplify the CSI as $h_{1,i}=\ldots=h_{N,i}=h^{(i)}$ for $i\in \mathcal{N}$, and the MSSV received is:
\begin{equation}\label{eq:fad_channel2}
\mathbf{z}_i^{\text{rev}}  =  h^{(i)} \sum_{k=1}^{N} \mathbf{z}_k + \mathbf{n}_i.
\end{equation}

In the uplink transmission scenario, as Fig.\ref{DeepMA}(c) shows, we can also simplify the CSI as $h_{i,1}=\ldots=h_{i,N}=h_i$ for $i\in \mathcal{N}$. In addition, the AWGN for all RMSSVs are the same, i.e., $\mathbf{n}_1 = \ldots=\mathbf{n}_N = \mathbf{n}$. Then we can know the MSSV received by the BS is:
\begin{equation}\label{eq:fad_channel3}
\mathbf{z}^{\text{rev}} = \sum_{i=1}^{N} h_i \mathbf{z}_i + \mathbf{n}.
\end{equation}

We note that, the SSV transmitting processes defined by equations \eqref{eq:fad_channel2} and \eqref{eq:fad_channel3} are actually two simplified cases of equation \eqref{eq:fad_channel1}. It also worth noting that in downlink and uplink scenarios, the BS needs to access the data contents before encoding or after decoding, and this requirement may cause privacy concerns for the UEs. In this paper, we assume that the BS can protect the privacy for the transmitted data, and we only focus on the multiple access problem in DeepMA.
}

{\color{black}
\subsection{Encoding Process}
In DeepMA, an encoder of an EDP is a deep learning model that conducts the JSCC and orthogonal signal modulation process, such that its input data can be transformed into a SSV that is suitable for wireless transmission with OMA.} More specifically, for EDP $i \in \mathcal{P}$, we denote the input data as $\mathbf{x}_i \in \mathbb{R}^L$, and before transmitting, it is encoded as a $K$-dimensional complex semantic feature data $\mathbf{y}_i \in \mathbb{C}^{K}$ by using its encoder model $E_{\phi_i}$, i.e.,
\begin{equation}
\mathbf{y}_i = E_{\phi_i}(\mathbf{x}_i, \gamma_i),
\end{equation}
where $\gamma_i$ denotes the channel SNR. We note that the transmission performance of an EDP is also influenced by the channel condition, and the channel SNR will also be an input data to the encoder. {\color{black}If the semantic feature $\mathbf{y}_i$ is not a vector, it will be reshaped as a vector stream for the following transmission process.} To meet the power constraint, $\mathbf{y}_i$ will be normalized by using the following equation:
\begin{equation}\label{eq:p_norm}
\mathbf{z}_i  = \sqrt{K P_z} \frac{\mathbf{y}_i}{\mathbf{y}_i^{*} \mathbf{y}_i},
\end{equation}
where $P_z>0$ denotes the average transmitting power constraint. {\color{black}In this paper, we call the above normalized semantic feature as the semantic symbol vector (SSV). The average power constraint $P_z$ can be set to a proper value that is larger than 0, for example, $P_z = 1$. After the power normalization, the SSVs are transmitted to the decoder sides.}

{\color{black}
In contrast to the SSVs obtained from the encoder in existing DeepJSCC models, in DeepMA we require the SSVs are mutually orthogonal, as given by:
\begin{equation}
\frac{1}{K} \mathbb{E} [\mathbf{z}_i^*\mathbf{z}_j] =
\begin{cases}
  P_z, & \mbox{if } i=j \\
  P_{\bot}, & \mbox{otherwise}.
\end{cases}
\end{equation}
In the equation above, the value of $P_{\bot}$ is very small and close to 0, i.e. $P_{\bot} \simeq 0$. The orthogonality above is learned during the training process of DeepMA, and it will be further illustrated and verified in the following subsections.
}

{\color{black}
\subsection{Decoding Process}
After the MSSVs are received, the DeepMA decoders conduct the joint orthogonal signal demodulation, and source-channel decoding processes, and finally obtain the reconstructed data. Before decoding, we must first compute its recovered MSSV (RMSSV) to reduce the impact of the wireless channel. For decoder $D_{\theta_i}$, its RMSSV is computed as follows:
\begin{equation}
\tilde{\mathbf{z}}_i = \frac{\mathbf{z}_i^{\text{rev}}}{h_{i,i}}.
\end{equation}
It is worth noting that in uplink scenario, the MSSV received of all decoders are the same, i.e., $\mathbf{z}_1^{\text{rev}} = \ldots = \mathbf{z}_N^{\text{rev}}=\mathbf{z}^{\text{rev}}$.

In a multi-access scenario, the RMSSV obtained may not contain the SSV transmitted from the paired encoder, and it is better to be abandoned because the decoding process will cause unnecessary computation cost, especially when the complexity of the decoder model is high. In addition, DeepMA itself has good data privacy protection ability and a decoder cannot recover the data transmitted from other unpaired encoders. We present examples to demonstrate this property in the following section.

\begin{figure}
\centering
\includegraphics[width=0.4\textwidth]{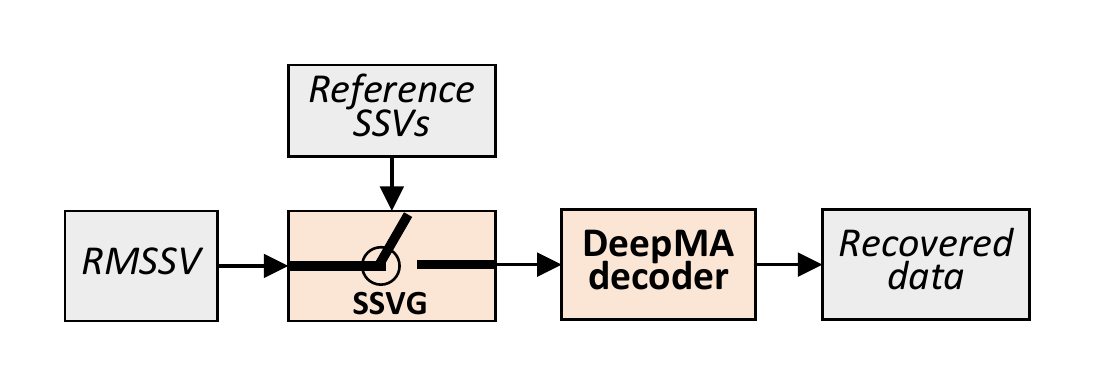}
\caption{{\color{black}User detection using a SSVG.}}
\label{fig:ssvg}
\end{figure}

To classify whether the RMSSV contains the desired SSV component, as Fig. \ref{fig:ssvg} shows, we use a SSV gate (SSVG) to conduct user detection operation by utilizing the orthogonality of the SSVs. More specifically, we first prepare a set of reference SSVs samples transmitted from the paired encoder at the decoder side. We denote the reference SSV samples as $\mathbf{z}_1^{\text{ht}},\ldots, \mathbf{z}_M^{\text{ht}}$, then we can compute the average absolute correlation degree (AACD), denoted as $R_i$, between the RMSSV and reference SSV samples, as given by:
\begin{equation}\label{eq:AACD}
R_i = \frac{1}{KM}\sum_{m=1}^{M} \left| \tilde{\mathbf{z}}_i^* \mathbf{z}_m^{\text{ht}} \right|.
\end{equation}
When the RMSSV contains a desired SSV component, $R_i$ will be a relatively large value, otherwise it will be a very small value. We note that the correlation between SSV and AWGN is nearly 0, thus AWGN will not influence the correlation value obtained by the equation above. In this way, we can use the thresholding operation to detect the target SSV. Let $R^{\text{th}}>0$ be the decision threshold, and we can know that, when $R_i\geq R^{\text{th}}$, then the RMSSV contains a target SSV, otherwise it is a non-desired one, and can be abandoned, as given by:
\begin{equation}
\text{decision} =
\begin{cases}
  \text{accept}, & \mbox{if }R_i\geq R^{\text{th}}\\
  \text{abandon}, & \mbox{otherwise}.
\end{cases}
\end{equation}

If RMSSV $\tilde{\mathbf{z}}_i$ is accepted, it will be used to recover the input data, as given by:
\begin{equation}
\tilde{\mathbf{x}}_i = D_{\theta_i}(\tilde{\mathbf{z}}_i, \gamma_i).
\end{equation}
In the decoding process above, the DeepMA decoder $D_{\theta_i}$ conducts joint orthogonal demodulation and source-channel decoding processes, such that the decoder can extract the target SSV and recover the input data from the RMSSV.
}

\subsection{DeepMA Training}
{\color{black}
As we have previously mentioned, DeepMA can be used in D2D, downlink, and uplink transmission scenarios to achieve co-frequency channel multiplexing. In the model training process, we recall that the transmission process defined by equations \eqref{eq:fad_channel2} and \eqref{eq:fad_channel3} are two simplified cases of equation \eqref{eq:fad_channel1}, and there is no need to train specific models for all the three scenarios because a DMANet trained in the D2D scenario can also be deployed in the uplink and downlink transmission scenarios.}

In DeepMA, each EDP only transmits and recovers its target data $\mathbf{x}_i$, and its loss function is defined as:
\begin{equation}
\mathcal{L}(\theta_i, \phi_i; \mathbf{x}_i) = d(\mathbf{x}_i, \tilde{\mathbf{x}}_i),
\end{equation}
where $d(\mathbf{x}_i, \tilde{\mathbf{x}}_i)$ denotes the distortion metric between the original input $\mathbf{x}_i$ and the recovered result $\tilde{\mathbf{x}}_i$. In this paper, we use the mean square error (MSE) as the distortion metric, and therefore we have $d(\mathbf{x}_i, \tilde{\mathbf{x}}_i) = \Vert \mathbf{x}_i - \tilde{\mathbf{x}}_i \Vert^2 / L$, and $L$ denotes the dimension of input data $\mathbf{x}_i$. It is worth noting that the final result $\tilde{\mathbf{x}}_i$ is recovered from an RMSSV that is mixed by the SSVs transmitted by all the DeepMA encoders, as formulated in equation \eqref{eq:fad_channel1}. The loss function of the whole DMANet is quite simple, and it is set to minimize the average distortions between the input data and the recovered results of all the EDPs, as given by:
\begin{equation}\label{eq:loss}
\begin{aligned}
\mathcal{L}(\Phi, \Theta; \mathbf{X}) & = \frac{1}{N}\mathcal{L}(\theta_i, \phi_i; \mathbf{x}_i) =\frac{1}{N}\sum_{i=1}^{N} d(\mathbf{x}_i, \tilde{\mathbf{x}}_i),
\end{aligned}
\end{equation}
where $\Theta = \left\{\theta_1,\ldots, \theta_N\right\}$ and $\Phi = \left\{\phi_1,\ldots, \phi_N\right\}$ denote the parameters of DeepMA encoders and decoders respectively, $\mathbf{X} = \left\{\mathbf{x}_1,\ldots,\mathbf{x}_N \right\}$ means the input training data samples. The loss function above enables us to train the decoders to recover their own target data from a received MSSV in wireless D2D transmission scenarios. By using a gradient descent algorithm, the parameters of the whole DMANet can be learned as:
\begin{equation}
 (\Phi^*, \Theta^*) = \text{argmin}_{\Phi, \Theta} \,\, \mathbb{E}_{\mathbf{x}} \left[\mathcal{L}(\Phi, \Theta; \mathbf{X})\right].
\end{equation}

We note that, except for source-channel coding/decoding, DeepMA encoders and decoders also will be trained with orthogonal modulation and demodulation processes, which are key enablers for achieving OMA-based multiple access with a shared physical channel. In the same way as existing DeepJSCC models, the proposed DeepMA is also trained in an end-to-end way. {\color{black}One main difference is that, since we want to achieve co-frequency channel multiplexing, in each training iteration the input data batch of each EDP must be different, such that the decoders can be trained with the ability to recover different data from an RMSSV. More specifically, in each training iteration, we need $N$ different data batches for the $N$ different EDPs. In this way, the decoder of one EDP will be trained to recover its own data from the received RMSSV, and ignore the impact of other SSV components transmitted from unpaired encoders.}

Algorithms 1 illustrate the process of one training iteration for DMANet-N in D2D transmission scenarios, in which $\mathcal{E}_{\Phi} = \left\{E_{\phi_1}, \ldots,E_{\phi_N}\right\}$ and $\mathcal{D}_{\Theta} = \left\{D_{\theta_1}, \ldots,D_{\theta_N}\right\}$ denote the DeepMA encoders and decoders respectively. In one training iteration, we first sample different data batches for different EDPs, then forward the data batches $\mathbf{x}_1, \ldots, \mathbf{x}_N$ through the DMANet, and compute the loss by using equation \eqref{eq:loss}, and further compute the gradients of all model parameters. {\color{black}For the channel model, the CSI and AWGN of each transmitted SSV are randomly generated such that the channel realizations are random.} Note that the user detection operation is not needed to be conducted in the training process. At last, the parameters $ (\Phi, \Theta)$ of DMANet can be updated by using the gradients obtained, and one training iteration is finished. We repeat the training iteration until the loss achieves convergence.

\begin{algorithm}
\caption{One training iteration for DMANet-$N$.}
\begin{algorithmic}[1]
\STATE  \textbf{Input}: {\color{black}Training dataset $\mathbf{X}$, transmitting power $P_z$, channel SNRs $\gamma_1, \ldots, \gamma_N$, learning rate $\eta$};
\STATE  \textbf{Output}: Model parameters $(\Theta, \Phi)$ of DMANet-N;
\STATE  \textbf{Transmitters}:
\STATE  \quad \textbf{for each DeepMA encoder $E_{\phi_i} \in \mathcal{E}_{\Phi}$}:
\STATE  \quad \quad Sample a different training data batch $\mathbf{x}_i$ from $\mathbf{X}$;
\STATE  \quad \quad Compute SSV $\mathbf{y}_i \leftarrow E_{\phi_i}(\mathbf{x}_i, {\color{black}\gamma_i})$;
\STATE  \quad \quad Power normalization $\mathbf{z}_i \leftarrow \sqrt{K P_z} \frac{\mathbf{y}_i}{\mathbf{y}_i^{*} \mathbf{y}_i}$;
\STATE  \quad \textbf{end for each}
\STATE  {\color{black}\textbf{Channel}:
\STATE  \quad Randomly generate CSI matrix $\mathbf{H}=[h_{i,j}]_{N\times N}$, in which $h_{i,j} \sim \mathcal{CN}(0,1)$;
\STATE  \quad \textbf{for each transceiver $i\in \mathcal{N}$}:
\STATE  \quad \quad Compute AWGN power $\sigma_i^2 = \frac{P_z}{10^{(\gamma_i/10)}}$;
\STATE  \quad \quad Randomly generate AWGN $\mathbf{n}_i \sim\mathcal{CN}(0, \sigma_i^2 \mathbf{I})$;
\STATE  \quad \textbf{end for each}}
\STATE  \textbf{Receivers}:
\STATE  \quad \textbf{for each DeepMA decoder $D_{\theta_i}\in \mathcal{D}_{\Theta}$}:
\STATE  \quad \quad Receive MSSV $\mathbf{z}_i^{\text{rev}}  = \sum_{k=1}^{N}  h_{k,i} \mathbf{z}_k + \mathbf{n}_i$;
\STATE  \quad \quad Compute RMSSV $\tilde{\mathbf{z}}_i \leftarrow {\mathbf{z}_i^{\text{rev}}}/{h_{i,i}}$;
\STATE  \quad \quad Recover data $\tilde{\mathbf{x}}_i \leftarrow D_{\theta_i}(\tilde{\mathbf{z}}_i, {\color{black}\gamma_i})$;
\STATE  \quad \textbf{end for each}
\STATE  Compute loss $\mathcal{L}(\Phi, \Theta; \mathbf{X})\leftarrow \frac{1}{N} \sum_{i=1}^{N} d(\mathbf{x}_i, \tilde{\mathbf{x}}_i)$;
\STATE  Update model parameters by gradient descent: $$\Theta \leftarrow \Theta - \eta \triangledown_{\Theta} \mathcal{L}(\Phi, \Theta; \mathbf{X}), \Phi \leftarrow \Phi - \eta \triangledown_{\Phi} \mathcal{L}(\Phi, \Theta; \mathbf{X}).$$
\end{algorithmic}
\end{algorithm}

\section{DeepMA for Wireless Image Transmission}

In this section, we evaluate the performance of the proposed DeepMA by conducting simulations on DeepMA-enabled wireless image transmission tasks. First, we describe the architecture of the DMANet used and the experimental settings. Next, we describe the SSV orthogonality to explain why DeepMA can achieve channel multiplexing and how to use SSV orthogonality to conduct user detection, at the same time showing its flexibility and privacy properties. Finally, we present the performance evaluation results, along with their analysis.

\begin{figure*}
\centering
\includegraphics[width=0.99\textwidth]{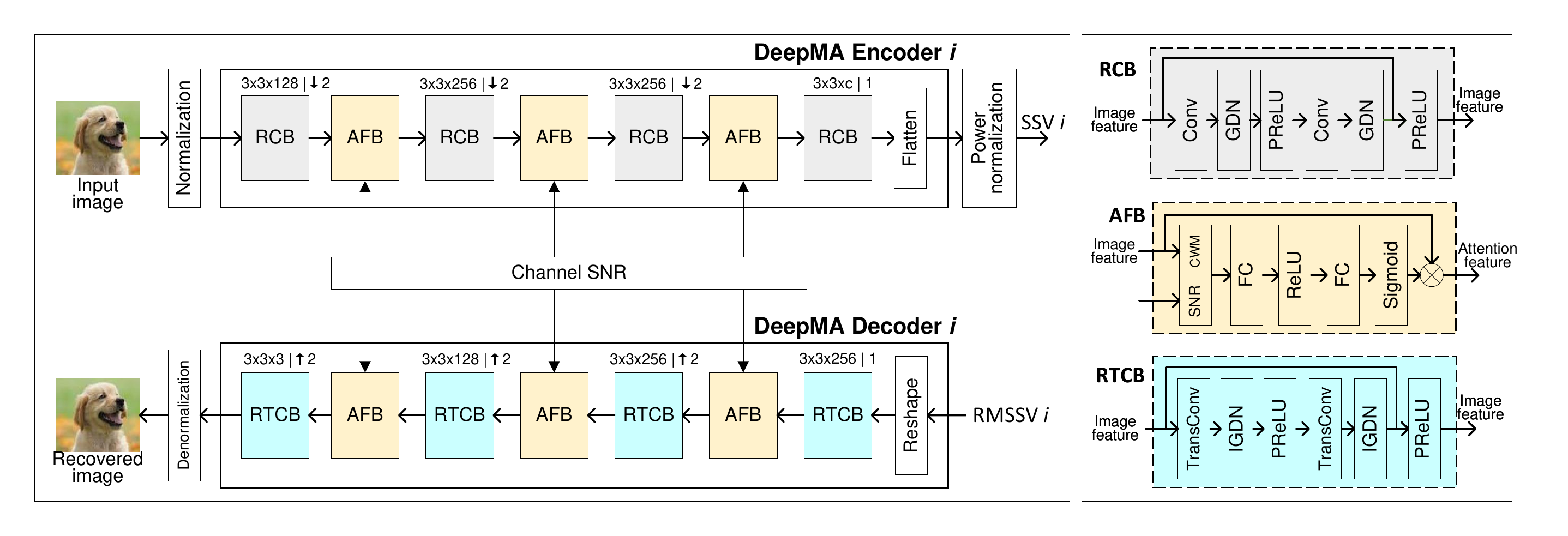}
\caption{Network architecture of one EDP of the DMANet used.}
\label{fig:network}
\end{figure*}

\subsection{Network Architecture}

Fig. \ref{fig:network} shows the architecture of the DMANet used for image transmission with channel multiplexing ability. First, we normalize the input image  with a value range [0, 1], and then passes it through the DeepMA encoder to obtain the SSV. The DeepMA encoder is used for conducting the joint source-channel coding and orthogonal signal modulation process, which is achieved by four residual convolutional blocks (RCBs) and three attention feature blocks (AFBs). The RCBs are used for extracting the image features by using convolutional operations, and the AFB is used for adjusting the image features obtained by the RCBs. In AFB, we first compute the channel-wise mean (CWM) values of the image features. By using the CWM and channel SNR information, we use a simple fully connection (FC) network to predict the importance weights of the feature channels and automatically adjust the learned features with the weights obtained. The authors of \cite{xu2021wireless} provide more detailed introduction of AFB. In each RCB, we use the generalized normalization transformation (GDN) as the normalization layer, which has been demonstrated effective in density modeling and image compression tasks \cite{balle2015density}\cite{balle2017end}. The RCB is designated as $m\times n \times c \,\,| \downarrow s$, where $m,n,c, s$ correspond to the height, width, output channel number, and downsampling stride length respectively of the convolutional layer used. When the stride length is $s=1$, then the shape of the output data is the same as the shape of the input data. When the stride length is $s=2$, the height and weight of the output data is half of the input data. For the channel number $c$, the maximum channel number is $256$, and in the first RCB, the channel number is $128$, which is helpful for reducing memory space when the input image is large. In the last RCB, the channel number $c$ is used for adjusting the data compression ratio, and a relatively larger value of $c$ produces a better image recovery quality, but the transmission efficiency will be reduced. We denote $H$ and $W$ as the height and width of the input image respectively because we have three downsampling operations with $s=2$. Then we know the shape of the output data of the last RCB which is ${H}/{8}\times{W}/{8} \times c$. In this way, we define the bandwidth efficiency performance as the following symbol per pixel (SPP) metric:
\begin{equation}
\text{SPP}(c) = \frac{{H}/{8}\times{W}/{8} \times c}{H \times W} = \frac{c}{64}.
\end{equation}
After the reshape operation, we transform the image feature as a $K$-dimensional complex SSV, and we can know $K = \frac{HW}{128}$. We note that a complex semantic symbol is composed of two real semantic features. Accordingly, we define complex SPP (CSPP) metric as follows:
\begin{equation}
\text{CSPP}(c) = \frac{c}{128}.
\end{equation}
We observe that $\text{SPP}(c) = 2\text{CSPP}(c)$. For DMANet-N, it can support at most $N$ transmissions simultaneously. Therefore the achievable minimal CSPP (MinSCPP) is:
\begin{equation}
\text{MinCSPP}(c, N) = \frac{c}{128N}.
\end{equation}
For example, if we set $c=128$, for an ADJSCC model or one EDP, the CSPP is 1, and for DMANet-2, the MinCSPP is 0.5. In this paper, when the MinCSPP of a DeepMA model is equal to the CSPP of an ADJSCC model, we say that their achievable code rates are the same.

The DeepMA decoder is used for conducting the joint signal filtering and channel-source decoding process to recover the original input image with the RMSSV obtained.
Like the DeepMA encoder, the architecture of a DeepMA decoder is composed of four residual transpose convolutional blocks (RTCB) and three AFBs. The main difference between RTCB and RCB is that RTCB uses a transconvolution operation for reconstructing the image and uses an inverse GDN (IGDN) operation for feature normalization. Accordingly, since we want to recover the original image from the RMSSVs received, in the last three RTCBs, we use three upsampling operations with stride length $s=2$ to ensure the size of the recovered image is the same as the original input image. Finally, we denormalize the output image as 8-bit integer pixels within the range [0, 255].

\subsection{Simulation Settings}
We evaluated the performances of our proposed DeepMA on CIFAR100 \cite{CIFAR} and Kodak24 \cite{Kodak24} data transmission tasks with channel multiplexing, and we also tested the performance of the state-of-the-art ADJSCC model for non-multiplexing transmissions.
For an EDP model and an ADJSCC model, when the model architecture and the achievable code rates are the same, we expect that ADJSCC will achieve better performance compared with the DeepMA model because achieving orthogonal modulation and filtering will incur additional model encoding and decoding resources, and the used model resource for joint source-channel coding will be reduced. Note that model resources in this paper means the black-box computation process of the model. We used the following simulation settings in our performance evaluation tests:

\textbf{CIFAR100 data transmission}: In this test, we will use CIFAR10 \cite{CIFAR10} data as the training data and validation data, and then test the transmission performance on CIFAR100 test data. Both CIFAR10 and CIFAR100 contain 60000 color image samples with size $32\times 32$, in which 5000 images are training samples, and the remaining 1000 images are test samples. We will first train the semantic communication models by using CIFAR10 training data, and at the end of each training epoch, we validate the performance of the current model using the CIFAR10 test data. We implement DMANet and the ADJSCC models in the Pytorch environment, the batch size per EDP is set as 64. The initial learning rate is set as $\eta = 5\times 10^{-4}$. After 100 epochs and 200 epochs, it is decreased as  $\eta = 1\times 10^{-4}$ and  $\eta = 5\times 10^{-5}$ respectively. The maximal number of training epochs is set as 400.
In each training epoch, we first update the model by training data, then evaluate the performance of the model on the validation data. If the achieved performance is better than the best performance previously achieved, we save the model as the final model, otherwise, the final model will not be updated. After training, we use CIFAR100 test data to test the transmission performance of the proposed DeepMA and ADJSCC models.

\textbf{Kodak24 data transmission}: In this test, we used the ImageNet \cite{deng2009imagenet} data to train the DeepMA and ADJSCC models pretrained on the CIFAR10 data. Since the model is previously trained on small-scale CIFAR10 data, the training time and the number of ImageNet training samples used can be reduced. The original ImageNet data has more than one million color images, and in this paper we only used ImageNet validation dataset as the training data and validation data, in which the first 45000 images are used as training samples and the remaining 5000 images are used as the validation dataset. The image size of ImageNet can be different, thus we set the input image size to $128\times 128$ by using the center-crop operation.
As pretrained models are used, the training epochs on ImageNet data are not required to be too large, and we set the maximal epoch number to 200. We set the initial learning rate to $\eta = 1\times 10^{-4}$, and then gradually decreasing to $5\times 10^{-5}$, $1\times 10^{-5}$, and $1\times 10^{-6}$ at the 100th, 160-th, and 190-th epochs respectively. In the same way with the CIFAR10 data training process, at the end of each training epoch, we tested the performance of the models obtained on the validation data, and if the performance is better than the best performance previously achieved, we updated the final model, otherwise, the final model remains unchanged.
After model training, we use the Kodak24 data to test the performance of the model obtained. The Kodak24 data is composed of 24 color images with size $768\times 512$, and it is a widely used benchmark for evaluating the image quality of image reconstruction tasks. Since the number of images is quite limited, we therefore used the average performance on 100 tests to avoid performance fluctuation caused by the uncertain SSV transmission process.

\subsection{Performance Evaluation Metrics}
To evaluate the quality of the images recovered, we used peak signal-to-noise ratio (PSNR) \cite{hore2010image} as the performance metric, and it is defined as:
\begin{equation}
\text{PSNR} = 10\log_{10} \frac{\text{MAX}^2}{\text{MSE}},
\end{equation}
where $\text{MSE}$ is the mean-squared error between the original input image and the image recovered, and $\text{MAX}$ means the allowable peak pixel value of the images. Since the images used is 8-bit per pixel, the signal peak value is $\text{MAX} = 2^8 -1=255$. A higher PSNR means better quality of the recovered image because the MSE becomes lower. In one DeepMA model, the PSNR performance results of the EDPs can be slightly different from each other. Thus, we used the average value of the PSNRs of all EDPs as the PSNR performance of a DMANet.

During the transmission process, the channel SNR may significantly affect the quality of the image recovered, and the channel SNR of EDP $i$ is defined as:
\begin{equation}
\gamma_i =  10\log_{10} \frac{P_z}{\sigma_i^2},
\end{equation}
where $P_z$ denotes the average transmitting power of the SSV $\mathbf{z}$, and as we have mentioned before we set $P_z=2P_0 = 2$. {\color{black}With a given channel SNR $\gamma_i$ and transmitting power $P_z$, we can compute the corresponding noise power $\sigma_i^2 = 10^{(-\gamma_i/10)}P_z$. Then, we can randomly generate the AWGN samples for each symbol of the MSSV received. With the randomly generated channel CSI matrix and AWGN noise samples, we can create random channels with given channel SNRs.}

\begin{figure*}
\centering
 \subfigure[Kodim22, $\mathbf{x}_1$, EDP $(E_{\phi_1}, D_{\theta_1})$]{
 \includegraphics[width=0.3\textwidth]{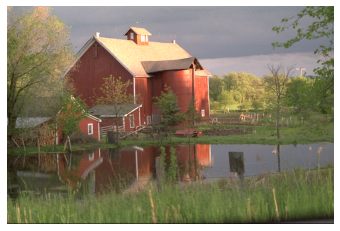}}
 \subfigure[Kodim23, $\mathbf{x}_2$, EDP $(E_{\phi_2}, D_{\theta_2})$]{
 \includegraphics[width=0.3\textwidth]{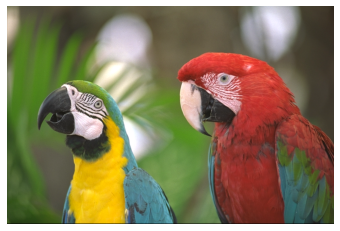}}
 \subfigure[Kodim24, $\mathbf{x}_3$, EDP $(E_{\phi_3}, D_{\theta_3})$]{
 \includegraphics[width=0.3\textwidth]{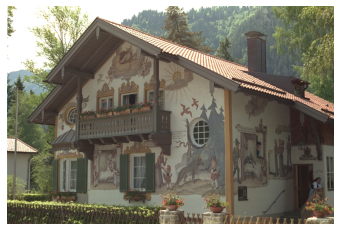}}
 \subfigure[$\mathbf{z}_1^{\text{rev}}  =\sum_{i=1}^{3} (h_{i,1}\mathbf{z}_i + \mathbf{n}_1)$, 33.53dB]{
 \includegraphics[width=0.3\textwidth]{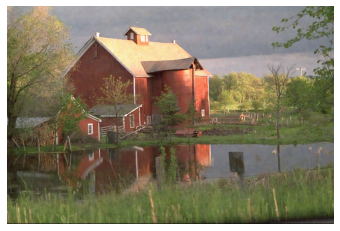}}
 \subfigure[$\mathbf{z}_2^{\text{rev}} =\sum_{i=1}^{3} (h_{i,2}\mathbf{z}_i + \mathbf{n}_2), D_{\theta_2}$, 36.47dB]{
 \includegraphics[width=0.3\textwidth]{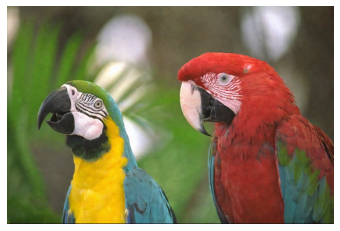}}
 \subfigure[$\mathbf{z}_3^{\text{rev}}=\sum_{i=1}^{3} (h_{i,3}\mathbf{z}_i + \mathbf{n}_3), D_{\theta_3}$, 35.01dB]{
 \includegraphics[width=0.3\textwidth]{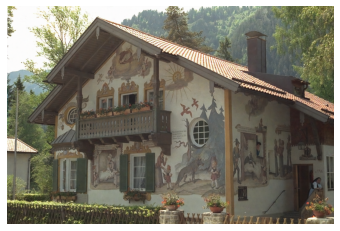}}
 \subfigure[$\mathbf{z}_1^{\text{rev}}=h_{1,1}\mathbf{z}_1 + \mathbf{n}_1, D_{\theta_1}$, 33.54dB]{
 \includegraphics[width=0.3\textwidth]{./Examples/kodak22_rec.png}}
 \subfigure[$\mathbf{z}_2^{\text{rev}}=h_{2,2}\mathbf{z}_2 + \mathbf{n}_2, D_{\theta_2}$, 36.35dB]{
 \includegraphics[width=0.3\textwidth]{./Examples/kodak23_rec.png}}
 \subfigure[$\mathbf{z}_3^{\text{rev}}=h_{3,3}\mathbf{z}_3 + \mathbf{n}_3, D_{\theta_3}$, 35.06dB]{
 \includegraphics[width=0.3\textwidth]{./Examples/kodak24_rec.png}}
 \subfigure[$\mathbf{z}_1^{\text{rev}}=\sum_{i=2}^{3} (h_{i,1}\mathbf{z}_i + \mathbf{n}_i), D_{\theta_1}$]{
 \includegraphics[width=0.3\textwidth]{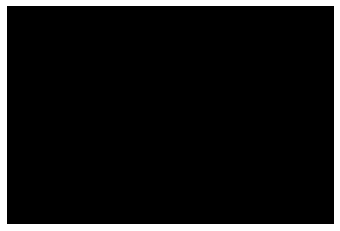}}
 \subfigure[$\mathbf{z}_2^{\text{rev}}=\sum_{i=1,i\neq 2}^{3} (h_{i,2}\mathbf{z}_i + \mathbf{n}_i), D_{\theta_2}$]{
 \includegraphics[width=0.3\textwidth]{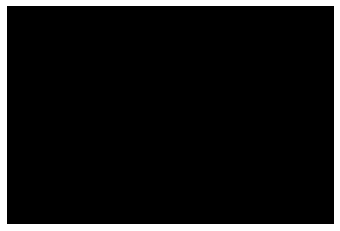}}
 \subfigure[$\mathbf{z}_3^{\text{rev}}=\sum_{i=1}^{2} (h_{i,3}\mathbf{z}_i + \mathbf{n}_i), D_{\theta_3}$]{
 \includegraphics[width=0.3\textwidth]{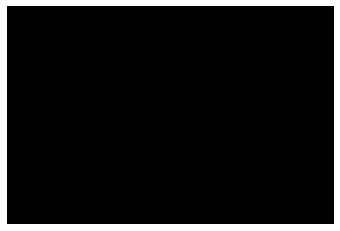}}
\caption{Examples of the images reconstructed by DMANet-3 for wireless image transmission with slow Rayleigh fading channel, we set MinCSPP to 0.5, and the channel SNRs of EDPs $(E_{\phi_1}, D_{\theta_1})$, $(E_{\phi_2}, D_{\theta_2})$ and $(E_{\phi_3}, D_{\theta_3})$ to 8 dB, 12 dB, and 16 dB respectively.}
\label{fig:sma3_examples}
\end{figure*}

\subsection{Illustrative Examples of DMANet-3}

In this subsection, we present some image examples reconstructed by DMANet-3 for the D2D wireless image transmission scenario. Fig. \ref{fig:sma3_examples} shows the example images, in which subfigures (a)-(c) are the three original input images, and subfigures (d)-(l) are the reconstructed images obtained with different RMSSVs. The three example images used are kodim22, kodim23, and kodim24, and they are denoted as $\mathbf{x}_1$, $\mathbf{x}_2$, and $\mathbf{x}_3$ respectively. We transmitted three images independently by three different transmitters, and the EDPs used for $\mathbf{x}_1$, $\mathbf{x}_2$, and $\mathbf{x}_3$ are $(E_{\phi_1}, D_{\theta_1})$, $(E_{\phi_2}, D_{\theta_2})$, and $(E_{\phi_3}, D_{\theta_3})$ respectively. The corresponding normalized SSVs of the three images are $\mathbf{z}_1$, $\mathbf{z}_2$, and $\mathbf{z}_3$. We set the channel to a slow Rayleigh fading channel, and we set the channel SNRs of the three paths to 8dB, 12dB, and 16dB. Using the settings above, we conducted the following three tests:

\textbf{Test 1: Channel multiplexing}. In this test, the wireless channel is multiplexed by the three transceivers, and the received MSSV of the receiver $k$ is $\mathbf{z}_k^{\text{rev}}  =\sum_{i=1}^{3} (h_{i,k}  \mathbf{z}_i + \mathbf{n}_k)$. We then can compute the RMSSV of $D_{\theta_1}$, $D_{\theta_2}$, and $D_{\theta_3}$ as: $\tilde{\mathbf{z}}_1 = \mathbf{z}^{\text{rev}}/h_1$, $\tilde{\mathbf{z}}_2 = \mathbf{z}^{\text{rev}}/h_2$, and $\tilde{\mathbf{z}}_3 = \mathbf{z}^{\text{rev}}/h_3$ respectively. Finally, we can obtain the final reconstructed images as: $\tilde{\mathbf{x}}_1 = D_{\theta_1}(\tilde{\mathbf{z}}_1)$, $\tilde{\mathbf{x}}_2 = D_{\theta_2}(\tilde{\mathbf{z}}_2)$, and $\tilde{\mathbf{x}}_3 = D_{\theta_3}(\tilde{\mathbf{z}}_3)$, and subfigures (d)-(f) plot the results obtained.
Visually, we can observe that the recovered images are the same as their original input images. More precisely, the PSNRs of three recovered images $\tilde{\mathbf{x}}_1$, $\tilde{\mathbf{x}}_2$, and $\tilde{\mathbf{x}}_3$ are {\color{black}35.53 dB}, 36.47 dB, and 35.01 dB respectively, which demonstrate that DMANet3 can achieve the channel multiplexing process while at the same time yields good image reconstruction quality.

\textbf{Test 2: Dedicated communication}. In this test, the channel is dedicated to only one EDP. In other words, only one transmitter transmits data at a one-time slot. In this scenario, the SSVs received from the receivers $D_{\theta_1}$, $D_{\theta_2}$, and $D_{\theta_3}$ are $\mathbf{z}_1^{\text{rev}}=h_1\mathbf{z}_1 + \mathbf{n}_1$, $\mathbf{z}_2^{\text{rev}}=h_2\mathbf{z}_2 + \mathbf{n}_2$, and $\mathbf{z}_3^{\text{rev}}=h_3\mathbf{z}_3 + \mathbf{n}_3$ respectively, and the subfigures (g)-(i) show the corresponding reconstructed images. We observe that the PSNRs of the reconstructed images $\tilde{\mathbf{x}}_1$, $\tilde{\mathbf{x}}_2$, and $\tilde{\mathbf{x}}_3$ are {\color{black}33.54 dB}, 36.35 dB, and 35.06 dB respectively, which show that DMANet can recover the results in a dedicated communication scenario. It is also worth noting that the quality of the images recovered is the same as in the previous channel multiplexing scenario. These results show that, in the channel multiplexing scenario, the undesired SSVs received by one EDP decoder do not affect the final reconstructed result. This is because these SSVs are mutually orthogonal, and the decoders can automatically filter out the non-desired SSVs. In this way, we note that DeepMA can work in a flexible way to support both multiplexing and non-multiplexing transmissions.

\textbf{Test 3: Cross decoding}. In this test, a DeepMA encoder does not send any SSV to the channel, but its paired DeepMA decoder tries to recover the image from the received MSSVs transmitted from other unpaired DeepMA encoders. For example, we try to recover the result of $\mathbf{z}^{\text{rev}}=\sum_{i=2}^{3} (h_i  \mathbf{z}_i + \mathbf{n}_i)$ by using the DeepMA decoder $D_{\theta_1}$, and subfigure (j) shows the result obtained. We observe that the image obtained is pure black, and all the pixels are 0. Subfigures (k)(l) show similar results. From these results, we conclude that a DeepMA decoder can only recover images from the MSSVs containing the SSV component transmitted by the paired DeepMA encoder, otherwise, nothing useful can be obtained. Therefore, a DeepMA model itself can therefore achieve good privacy protection.

\subsection{SSV Orthogonality}

\begin{table}
\centering
\caption{Correlation matrix of the three complex SSVs}
\begin{tabular}{|c|c|c|c|} \hline
   & $\mathbf{z}_1$ & $\mathbf{z}_2$ & $\mathbf{z}_2$ \\ \hline
  $\mathbf{z}_1$ & 2 & $2.32\times 10^{-5}$ & $2.04 \times 10^{-5}$ \\ \hline
  $\mathbf{z}_2$ & $2.32\times 10^{-5}$ & 2 & $9.55 \times 10^{-5}$ \\ \hline
  $\mathbf{z}_3$ & $2.04 \times 10^{-5}$ & $9.55 \times 10^{-5}$ & 2 \\ \hline
\end{tabular}
\label{tab:Rz}%
\end{table}%

\begin{table}
\centering
\caption{Correlation matrix of the three real SSVs}
\begin{tabular}{|c|c|c|c|} \hline
   & $\mathbf{v}_1$ & $\mathbf{v}_2$ & $\mathbf{v}_2$ \\ \hline
  $\mathbf{v}_1$ & 1 & $-8.01\times 10^{-6}$ & $1.02 \times 10^{-5}$ \\ \hline
  $\mathbf{v}_2$ & $-8.01\times 10^{-6}$ & 1 & $-4.77\times 10^{-5}$ \\ \hline
  $\mathbf{v}_3$ & $1.02\times 10^{-5}$ & $-4.77 \times 10^{-5}$ & 1 \\ \hline
\end{tabular}
\label{tab:Rv}%
\end{table}%

We have seen that DeepMA achieves a flexible, effective, and secure communication process with channel multiplexing. The reason why DeepMA can be so powerful is because the DeepMA model can achieve an orthogonal signal modulation process, such that the recovered results of one decoder will not be influenced by other SSVs sent from the unpaired encoders. In this subsection, we provide an example to demonstrate the orthogonality of the SSVs. In this test, we used the environment settings as the test used in subfigures \ref{fig:sma3_examples}(d)(e)(f). To quantify the orthogonality, we first define the correlation metric between complex SSVs $\mathbf{z}_i$ and $\mathbf{z}_j$ as follows:
\begin{equation}\label{eq:R_z}
R_z(\mathbf{z}_i, \mathbf{z}_j) = \frac{1}{K} \mathbb{E} [\mathbf{z}_i^*\mathbf{z}_j],
\end{equation}
We can see that when $i=j$, the correlation metric above is equal to the average signal power of the SSV, i.e., $R(\mathbf{z}_i, \mathbf{z}_j) = P_z$. As we have mentioned before, in this paper, we set $P_z = 2$.
We compute the correlation matrix of the SSVs obtained from the DMANet3 on kodim22, kodim23, and kodim24, and Table \ref{tab:Rz} presents the results. We can observe that, for $i,j \in \left\{1,2,3\right\}$, when $i\neq j$, the value of $R(\mathbf{z}_i, \mathbf{z}_j)$ are very close to zero, which means that the three SSVs $\mathbf{z}_1$, $\mathbf{z}_2$, and $\mathbf{z}_3$ are orthogonal to each other.
Next, we transform the $K$-dimensional complex SSVs as $2K$-dimensional real SSVs as follows: $\mathbf{v}_i = [\text{Re}(\mathbf{z}_i); \text{Im}(\mathbf{z}_i)]$, then we can define the correlation metric between real SSVs $\mathbf{v}_i$ and $\mathbf{v}_j$ as follows:
\begin{equation}\label{eq:R_v}
R_v(\mathbf{v}_i, \mathbf{v}_j) = \frac{1}{2K} \langle\mathbf{v}_i,\mathbf{v}_j \rangle,
\end{equation}
where $\langle\mathbf{v}_i,\mathbf{v}_j \rangle$ means the inner product of $\mathbf{v}_i$ and $\mathbf{v}_j$. According to the above definition, we can expect that $R_z(\mathbf{z}_i, \mathbf{z}_j)=2R_v(\mathbf{v}_i, \mathbf{v}_j)$. Accordingly, we compute the correlation matrix of the three real SSVs, and Table \ref{tab:Rv} shows the result obtained. Again, we observe that the three real SSVs $\mathbf{v}_1$, $\mathbf{v}_2$, and $\mathbf{v}_3$ are orthogonal to each other. We note that the SSVs are not strictly orthogonal to each other, as we can see the average inner product of two different real SSVs is only close to zero, but not equal to zero.

\begin{figure}
\centering
\includegraphics[width=0.45\textwidth]{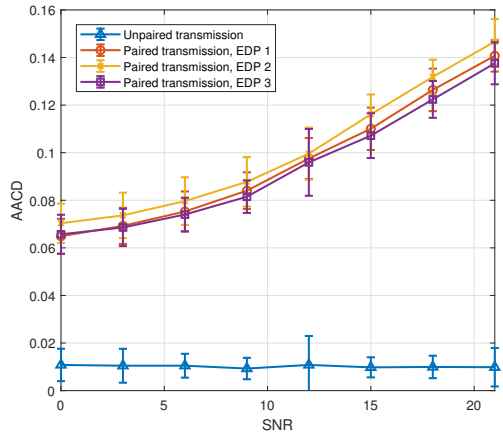}
\caption{{\color{black}AACD examples in user detection.}}
\label{fig:aacd}
\end{figure}

{\color{black}
The results above show the mutual orthogonality of the transmitted SSVs. Next we show how to use AACD defined in equation \eqref{eq:AACD} to conduct the user detection process. Recall that the AACD quantifies the correlation between the RMSSV received and reference SSVs.
Again, we use the same three images as the examples and test the AACDs with increasing channels SNRs in different scenarios. In each test, when transmitting one image, the SSVs of the other two images are used as the references for computing the AACD. Since the channel uncertainty has an impact on the AACD values, we get results for 100 repeats. Fig. \ref{fig:aacd} shows the results obtained, and we have the following observations:
\begin{itemize}
  \item \textbf{Unpaired transmission}: In this test, the decoder of one EDP receives a MSSV data transmitted from other two unpaired encoders, and we observe that the AACD distributions remains stable with a mean value 0.0103 and a standard deviation 0.0070, which are very small.
  \item \textbf{Paired transmission}: In this test, we only has one EDP transmit data, and we observe that, due to the impact of AWGN, the AACD values increases  when the channel’s SNR increases. In a very low SNR regime (e.g., 0dB), the mean value and standard deviation are about 0.067 and 0.0083 respectively, which are quite distinct compared with the unpaired transmission scenario. We observe that, by using only two reference SSVs, we can achieve an effective user detection process.
\end{itemize}

From the results above, we note that joint source-channel encoding and orthogonal modulation processes are conducted in the DeepMA encoder, and the SSVs obtained are orthogonal to each other, which enable different transceivers to transmit data over a shared physical channel without causing mutual interferences. In addition, as we have mentioned before, by using the SSV orthogonality, we can use the AACD defined in equation \eqref{eq:AACD} for solving the user detection problem in DeepMA, and then only execute the decoding process when the paired encoder transmits data.
}

\begin{figure*}
\centering
 \subfigure[]{
 \includegraphics[width=0.4\textwidth]{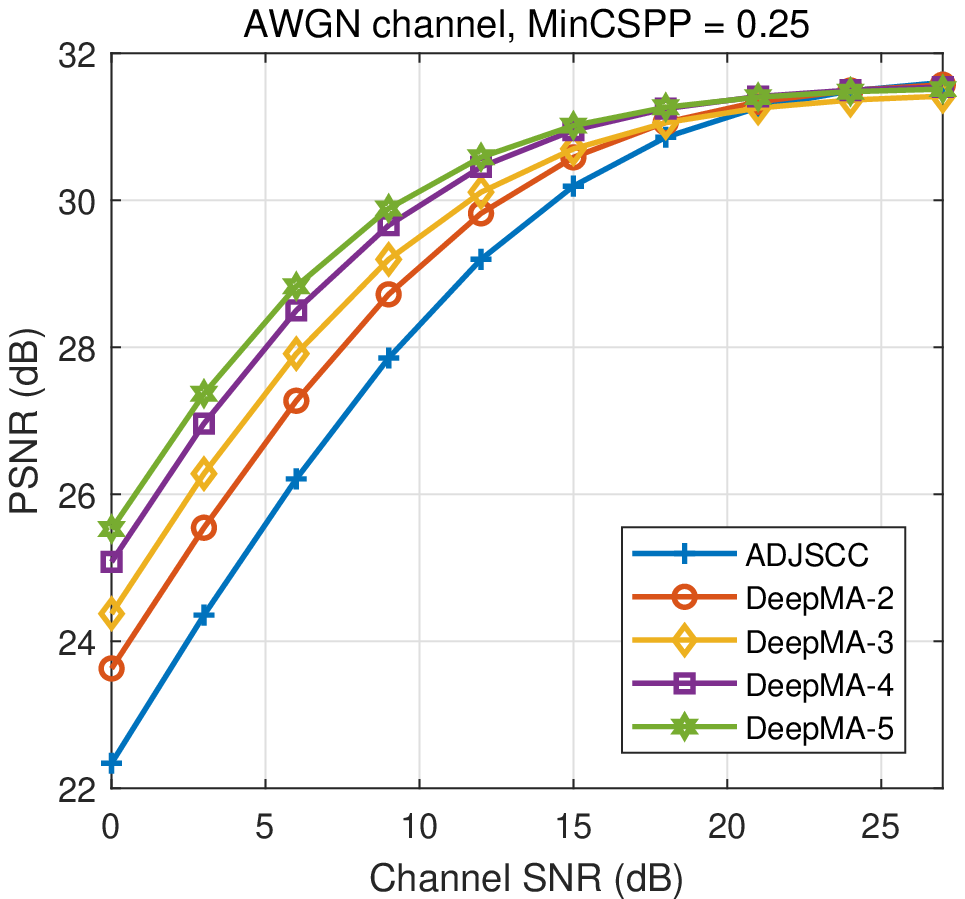}}
  \subfigure[]{
 \includegraphics[width=0.4\textwidth]{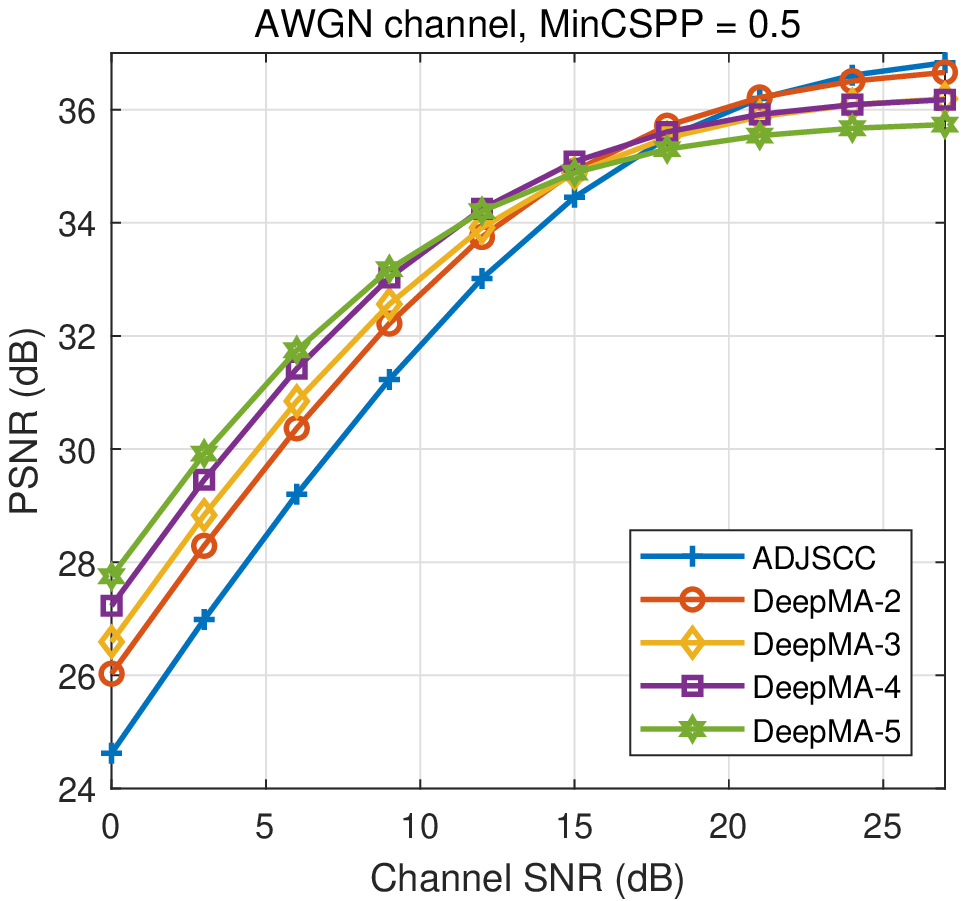}}
\caption{Comparison of the PSNR performances of DeepMA and ADJSCC on CIFAR100 data in the AWGN channel case.}
\label{fig:cifar_awgn}
\end{figure*}

\begin{figure*}
\centering
 \subfigure[]{
 \includegraphics[width=0.4\textwidth]{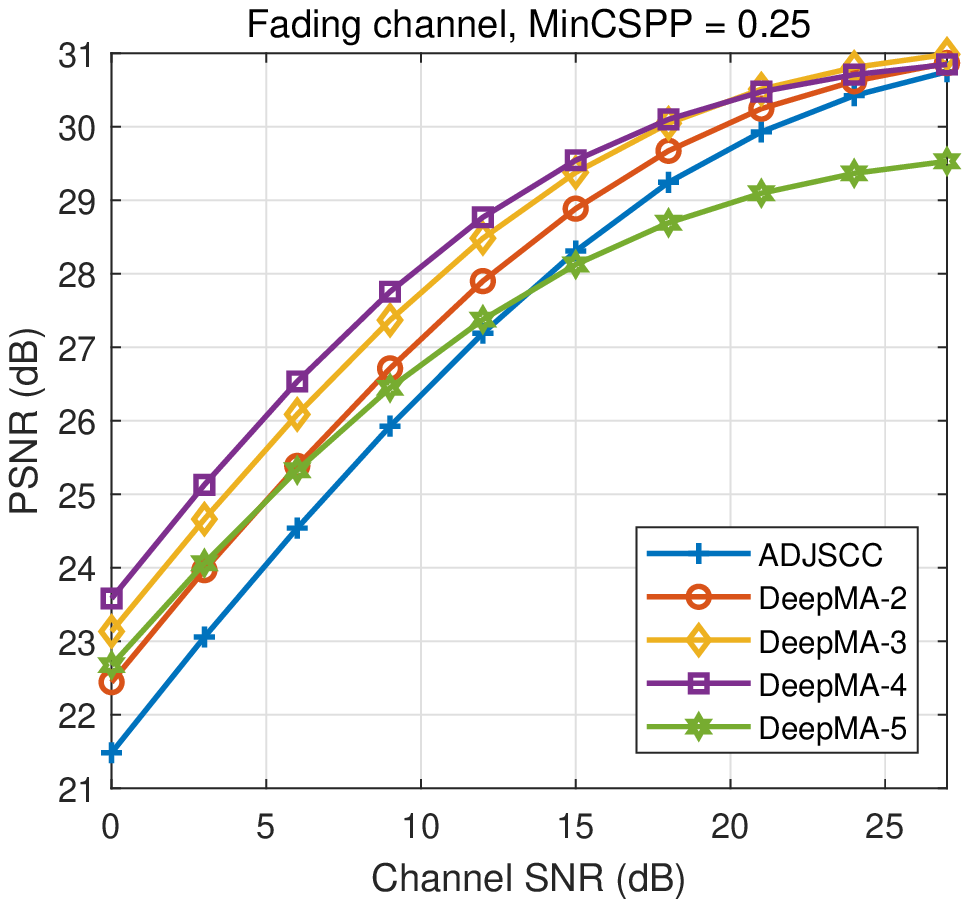}}
  \subfigure[]{
 \includegraphics[width=0.4\textwidth]{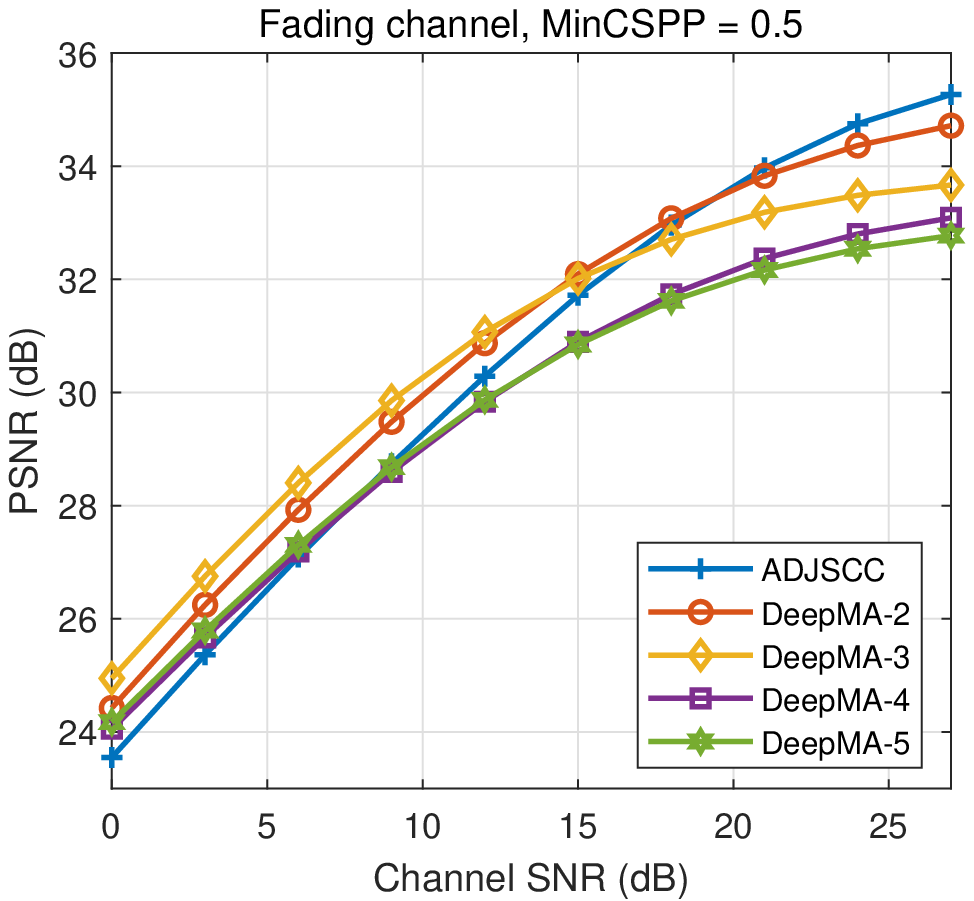}}
\caption{Comparison of the PSNR performances of DeepMA and ADJSCC on CIFAR100 data in the slow Rayleigh fading channel case.}
\label{fig:cifar_fading}
\end{figure*}

\subsection{Results on CIAFR100 Data}

In this subsection, we evaluate the PSNR performances of the proposed DeepMA and ADJSCC models on CIFAR100 data transmissions. Our main goal is to demonstrate that the proposed DeepMA can achieve effective channel multiplexing in wireless transmission scenarios, and at the same time show how the number of EDPs influences the performance of a DeepMA model. In a general D2D communication scenario, the SNRs of the transceivers can be different. To control the channel SNR and improve the legibility of the results, we set the SNRs of all transceivers to be the same. As aforementioned, downlink and uplink transmissions are two simplified of D2D transmissions, and we will test the performances of the D2D transmission case, and results of uplink and downlink transmission will not be presented due to the space limit.

Figs. \ref{fig:cifar_awgn} and \ref{fig:cifar_fading} show the PSNR performance results obtained with increasing channel SNRs. In all tests, the network architecture of the ADJSCC model used is the same as the EDP models of the DMANets used. For the ADJSCC model, we evaluated the PSNR performances with two code rates, namely CSPP=0.25 and CSPP=0.5, and the corresponding output channels of the encoder model are 32 and 64. For a DMANet-$N$ model, we also have $32N$ and $64N$. It is worth noting that, for one EDP, its CSPP is $N$ times the MinCSPP of DMANet-$N$, i.e., CSPP=$N\times$MinCSPP. We denote the two code choices as $\alpha=\left\{0.25, 0.5\right\}$, and from the two figures, we {\color{black}observe} that, when the bandwidth efficiencies are the same, the proposed DMANet can even achieve higher PSNRs when the channel SNR is not very high, but the PSNR performances of ADJSCC can be higher when channel SNR is high. These results first demonstrate that the proposed DeepMA method can achieve an effective OMA process and a stronger transmission ability than ADJSCC models for point-to-point transmission in a low SNR regime. One plausible reason is because when the bandwidth efficiency of ADJSCC and DMANet-$N$ are the same, the code length of DMANet-$N$ is $N$ times than the SSV of ADJSCC, and therefore incurs more total code resources on the channel coding process. In addition, in the training process, an EDP is trained with the ability to recover data from the interferences of other EDPs, which may be helpful for further enhancing the noise-resiliency capability of the EDPs. When the channel SNR is very high (e.g., 25 dB), the impact of noises is quite limited, and the code resources will be mainly used for the source coding process. However, the DMANet also needs to allocate more resources to the orthogonal modulation process, and therefore its remaining source coding resources will be lower than ADJSCC, and the PSNR performance achieved will be relatively lower.

Next, we compare the PSNRs achieved by the DMANet models with different numbers of EDPs. We denote the $\text{PSNR}^{(N)}$ as the PSNR performance of the DMANet-$N$ model, and let $\text{PSNR}^{(1)}$ be the PSNR of the ADJSCC model. Then, from Fig. \ref{fig:cifar_awgn}, we can see that, in the low SNR regime, the PSNR performance rank is $\text{PSNR}^{(1)}>\text{PSNR}^{(2)}>\ldots>\text{PSNR}^{(5)}$, while in the high SNR regime, the PSNR performance rank is $\text{PSNR}^{(1)}<\text{PSNR}^{(2)}<\ldots<\text{PSNR}^{(5)}$. Possible reasons could include: with the increasing EDN number $N$, the code length of a DAMNet model becomes larger, and it is trained with a stronger ability for dealing with more complex interferences. It also can be possible that the DMANet model has more total channel coding resources which increase the noise-resiliency ability. When the channel SNR is high, the average source coding resources reduce because model resources for orthogonal modulation increase, and the source coding performance decreases. The results in Fig. \ref{fig:cifar_fading} are the same as those in Fig. \ref{fig:cifar_awgn} when the EDP number is smaller than 4. We observe that, in Fig. \ref{fig:cifar_fading}(a), the performance of DeepMA-$5$ is significantly lower than ADJSCC and other DeepMA schemes. Similarly, in Fig. \ref{fig:cifar_fading}(b), the performance of DeepMA-$4$ and DeepMA-$5$ are lower than the other schemes. This result again shows that achieving the orthogonal modulation and demodulation process requires non-negligible model resources, especially as the EDP number increases. From these results, we can conclude that, in a complex fading channel case, a too-large EDP number may cause a negative impact on the transmission performance, and it is not suggested to use a too large EDP number to avoid significant performance loss.

\begin{figure*}
\centering
 \subfigure[]{
 \includegraphics[width=0.4\textwidth]{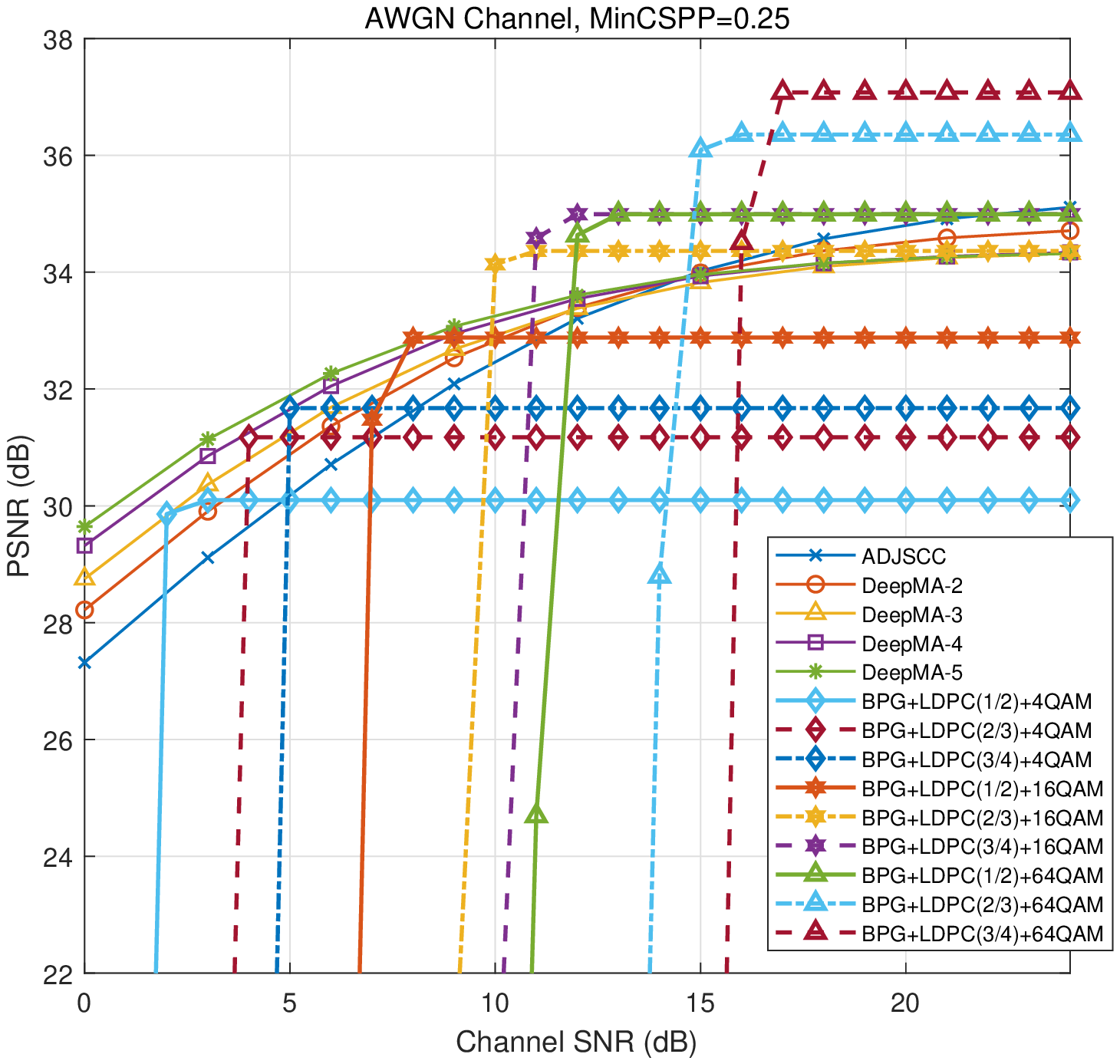}}
  \subfigure[]{
 \includegraphics[width=0.4\textwidth]{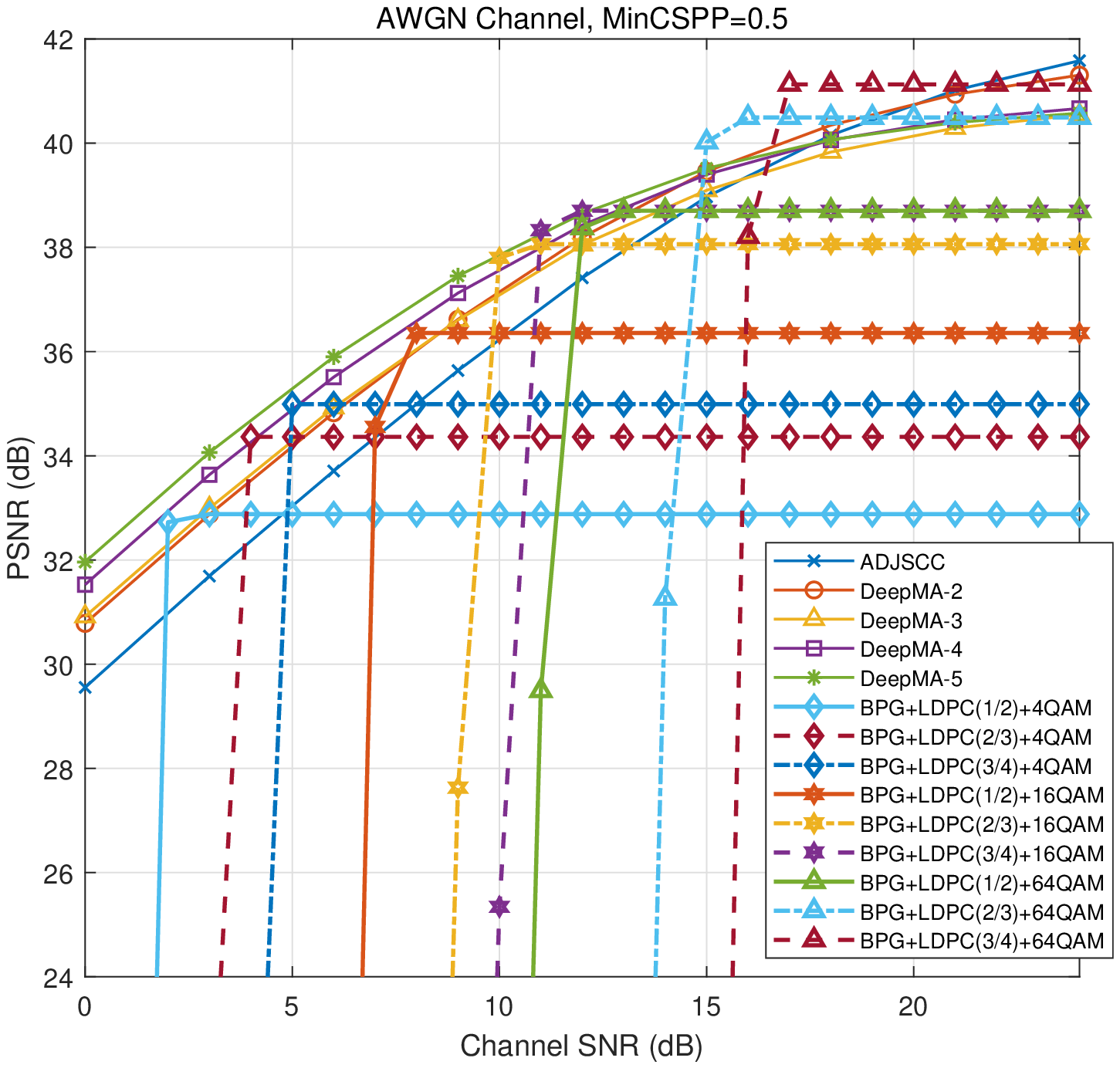}}
\caption{{\color{black}Comparison of the PSNR performances of DeepMA, ADJSCC, and conventional BPG+LDPC-based transmission schemes achieved on the  Kodak24 data in the AWGN channel case.}}
\label{fig:kodark24_comp_awgn}
\end{figure*}

\begin{figure*}
\centering
 \subfigure[]{
 \includegraphics[width=0.4\textwidth]{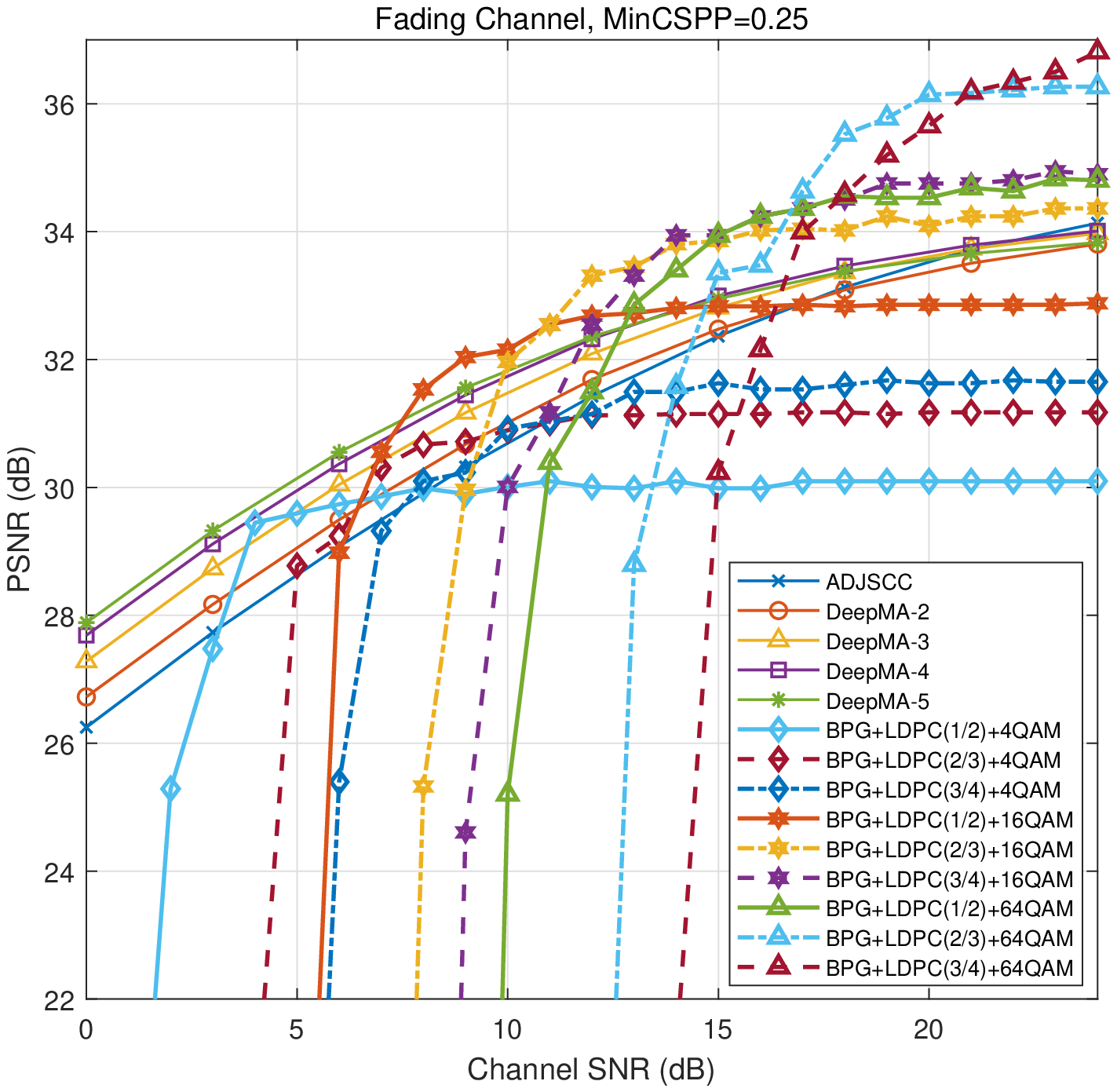}}
  \subfigure[]{
 \includegraphics[width=0.4\textwidth]{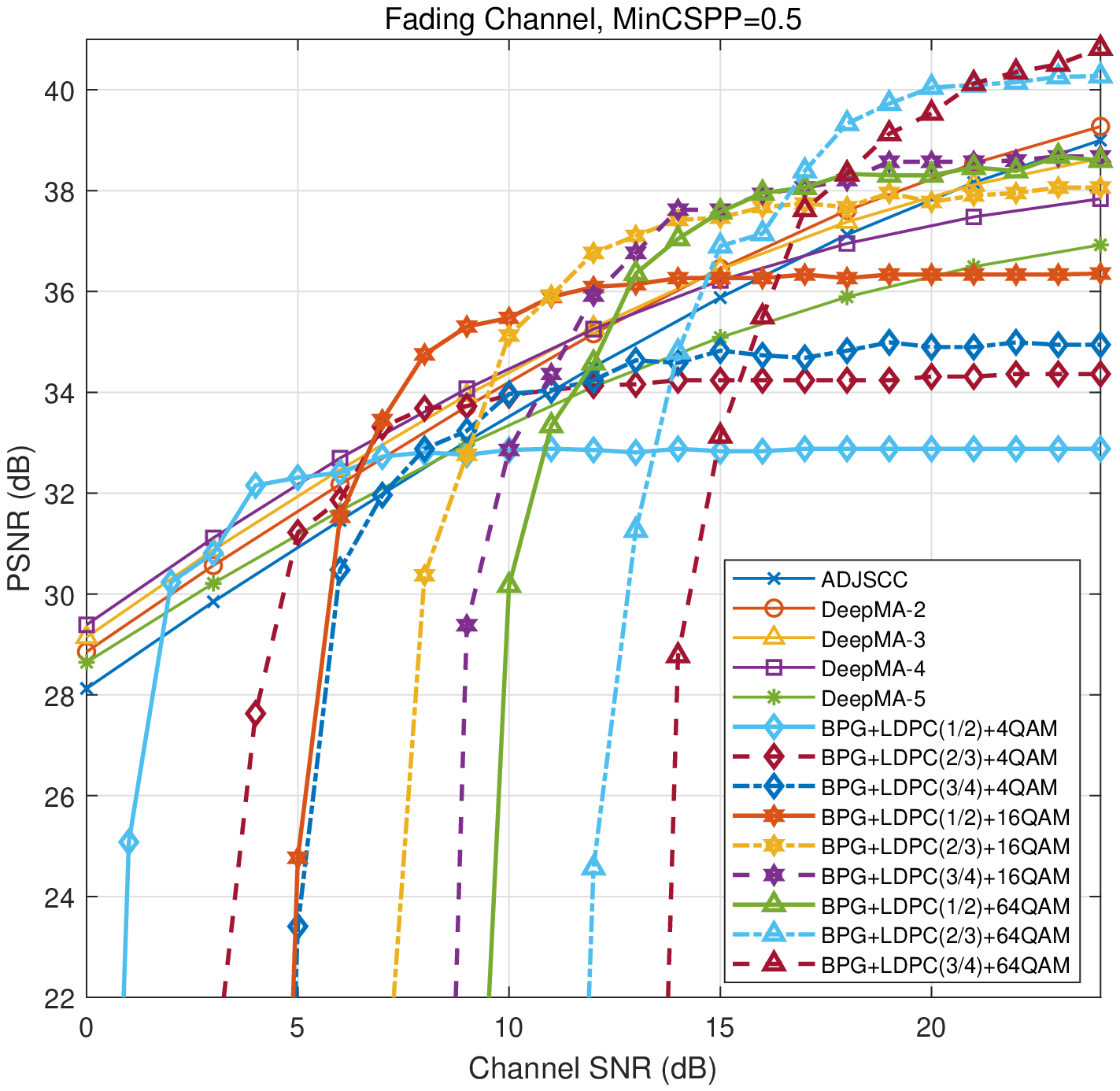}}
\caption{{\color{black}Comparison of the PSNR performances of DeepMA, ADJSCC, and conventional BPG+LDPC-based transmission schemes achieved on the Kodak24 data in the slow Rayleigh fading channel case.}}
\label{fig:kodark24_comp_fading}
\end{figure*}

\subsection{Results on Kodak24 Data}

In this subsection, we compare the PSNR performances of ADJSCC and the proposed DeepMA method achieved on the kodak24 data, whose resolution is much higher than the CIFAR100 data. For a fair comparison, in each test, the architecture of a used ADJSCC model is the same with one EDP of a DeepMA model, and the CSPP of the ADJSCC model is equal to the MinCSPP of the DeepMA model. Except for ADJSCC and DeepMA models, we also evaluated the performance of conventional wireless communications schemes that use BPG and LDPC codecs to achieve source coding and channel coding respectively. BPG is a well-known image \cite{BPG} codec based on discrete Fourier transform and intra-frame encoding, and it can achieve better image compression performance compared with the JPEG and the JPEG2000 standard \cite{BPG}. For LDPC, we used the channel coding scheme used in IEEE 802.11n protocol \cite{mahdi2020multirate}. We set the code block length to 1296, and we evaluated the performance of the following three code combinations $(648, 648)$, $(864, 432)$, and $(972, 324)$, and the corresponding code rates are 1/2, 2/3, {\color{black}and 3/4} respectively. In general, a relatively larger code rate increases the transmission efficiency but decreases the noise resiliency capability.  For the conventional communications scheme, the term $\text{LDPC}(R)$ means that the code rate of the LDPC code is $R$, and $R\in \left\{1/2,2/3,3/4\right\}$. For signal modulation, we tested the following 3 modulation orders: 4QAM, 16QAM, and 64QAM, and the corresponding number of bits per symbol is 2, 4, and 6 respectively. We note that the modulation order also significantly influences transmission efficiency and reliability. A relatively larger modulation order means a higher transmission efficiency and a lower transmission reliability. For conventional multiple access schemes, we assume that the OFDMA and TDMA schemes are used, and we equally allocate the frequency or timeslot resources to the UEs, such that the code length of each UE in the transmission is the same as each other. The proposed DeepMA can support co-frequency transmission, while OFDMA and TDMA cannot do so because they must allocate dedicated transmission resources for each UE. We did not compare the performance to CDMA-based multiple access schemes, because CDMA uses orthogonal basis codes to represent 1-bit data, and this digital coding process will greatly reduce the transmission bandwidth efficiency. Given the same bandwidth efficiency, the PSNR performances achieved with CDMA-based multiple access will be much lower than OFDMA and TDMA.

Figs. \ref{fig:kodark24_comp_awgn} and \ref{fig:kodark24_comp_fading} show the results obtained in the case of the AWGN channel and the slow Rayleigh fading channel cases. In a general D2D communication scenario, the SNRs of the transceivers can be different. Similar to the results in Figs. \ref{fig:cifar_awgn} and \ref{fig:cifar_fading}, we can observe that, compared with ADJSCC, in both AWGN and slow Rayleigh fading channels, the proposed DeepMA can achieve higher PSNR performances in the low SNR regime (e.g., 5dB in Fig. \ref{fig:kodark24_comp_awgn}(a)), but lower PSNR performances in a high SNR regime (e.g., 24dB in Fig. \ref{fig:kodark24_comp_awgn}(b)). In addition, we can also see that, in the slow Rayleigh fading channel case, a high EDP number will decrease the PSNR performance. We have explained the reasons for these results in the previous subsection, and we do not repeat the explanations here.

Finally, we compare the PSNR performances of the proposed DeepMA, ADJSCC, and conventional BPG+LDPC-based transmission schemes. We observe from, Figs. \ref{fig:kodark24_comp_awgn} and \ref{fig:kodark24_comp_fading}, that the performances of conventional schemes suffered from the cliff effect. Moreover, when the channel SNR is very low, e.g., lower than 0dB, the conventional schemes used even fail to transmit data. On the contrary, for both ADJSCC and the proposed DeepMA, the cliff effect does not exist, and they can still transmit data even when the channel SNR is very low. These results highlight the noise-resilience ability of ADJSCC and DeepMA. However, except for a very low SNR regime, the results obtained show that both ADJSCC and the proposed DeepMA cannot achieve better PSNR performances compared with conventional schemes. When MinCSPP is 0.25, conventional LDPC+BPG schemes with 64QAM can achieve much higher PSNR than DeepMA. This performance gap becomes much smaller when the MinCSPP is 0.5, and we note that DeepMA can achieve closer performance to conventional schemes. These results show that DeepMA still needs more effective deep learning models and training methods to improve its ability. However, in this work, our main goal is to demonstrate the effectiveness of DeepMA rather than achieving higher transmission performance than conventional BPG+LDPC schemes in all situations. In future work, we can explore the design of a more powerful network architecture to improve the transmission performance of the proposed DeepMA model.

\section{Conclusion}

In this paper, we have proposed DeepMA for achieving OMA-based channel multiplexing in semantic communication systems.
We formulated the DeepMA models for D2D, downlink, and uplink wireless communication scenarios, and we illustrated their training algorithm. We demonstrated the performance of the proposed DeepMA model on wireless image transmission tasks, and we presented the network architecture of the models used. We trained the ADJSCC and DeepMA models by using both low-resolution and small-scale CIFAR10 data and high-resolution and large-scale ImagNet data, and we evaluated the PSNR performances by using the CIFAR100 data and the kodak24 data. The results obtained show that the proposed DeepMA method can achieve effective, flexible, and privacy-preserved multiple access, and also showed that achieved orthogonal signal modulation will reduce performance compared with an ADJSCC model with the same encoding and decoding resources.

{\color{black}
\section*{Acknowledgment}
We thank the anonymous reviewers for their valuable comments which helped us improve the content and presentation of this paper.
}

\bibliographystyle{unsrt}
\bibliography{References}


\end{document}